\documentclass[aps,prb]{revtex4}

\usepackage{graphicx}

\begin{document}


\title{Epitaxial growth in dislocation-free strained alloy films: 
Morphological and compositional instabilities}

\author{Zhi-Feng Huang}
 \email{zfh@physics.utoronto.ca}
\author{Rashmi C. Desai}
 \email{desai@physics.utoronto.ca}
 \homepage{http://www.physics.utoronto.ca/people/faculty/desai.html}
\affiliation{%
Department of Physics, University of Toronto, \\
Toronto, Ontario, Canada M5S 1A7
}%


\begin{abstract}
The mechanisms of stability or instability in the strained alloy 
film growth are of intense current interest to both theorists and 
experimentalists. We consider dislocation-free, coherent, growing 
alloy films which could exhibit a morphological instability without 
nucleation. We investigate such strained films by developing a 
nonequilibrium, continuum model and by performing a linear 
stability analysis. The couplings of film-substrate misfit strain, 
compositional stress, deposition rate, and growth temperature 
determine the stability of film morphology as well as the surface 
spinodal decomposition. We consider some realistic factors of 
epitaxial growth, in particular the composition dependence of 
elastic moduli and the coupling between top surface and underlying 
bulk of the film. The interplay of these factors leads to new 
stability results. In addition to the stability diagrams both 
above and below the coherent spinodal temperature, we also 
calculate the kinetic critical thickness for the onset of instability 
as well as its scaling behavior with respect to misfit strain and 
deposition rate. We apply our results to some real growth systems 
and discuss the implications related to some recent experimental 
observations.
\end{abstract}

\pacs{81.15.Hi, 64.75.+g, 68.55.Jk}
\maketitle

\section{Introduction}

The techniques and technology related to fabricating large single 
crystals of elemental materials like silicon (Si) or compounds 
like gallium arsenide (GaAs) has been perfected for decades now. 
Polished wafers of such single crystalline semiconductor materials 
are now routinely used as substrates, on which a series of thin 
layers are grown to create a desired heterostructure. Thin films 
of increasing complexity are now being fabricated due to 
constraints on operating conditions and performance of electronic 
and optical devices.

When the growth of the superimposed material layers is based on
the atomic pattern of the substrate underneath, as if it is the 
extension of substrate structure, then the process is referred to 
as an epitaxial growth. Heteroepitaxy is the technique in which 
a thin film of a different material is grown on a substrate 
(as for example in a layer growth of AlAs on GaAs), and in which 
the deposited layer has the same crystal orientation as the 
substrate. Many materials (especially III-V compounds), though 
composed of different atoms, have identical crystal structures 
and nearly identical lattice spacing. A match of lattice spacing 
is important in epitaxial growth since it can minimize and possibly 
eliminate local strains in the growing film, and thus produce 
good quality thin solid layers. For example, one can in principle
lattice-match an InGaAsP quaternary alloy of appropriate composition 
to either InP or GaAs, and tailor-make a semiconductor film which 
will have the electronic band gap within the range between $0.8$ eV 
and $1.7$ eV.

Although the alloy composition can be tuned to obtain a desirable band 
gap, the resulting device is useful only if the alloy composition remains
homogeneous. In a typical epitaxial growth technique like molecular beam
epitaxy (MBE), the stability of the growing film can depend on many
variables. The homogeneity of the growing film clearly depends on the 
surface mobility of the material being deposited, the substrate 
temperature $T$, and the deposition rate $v$. 
Further, different sized atoms that are being deposited to fabricate an
alloy film induce stresses in the substrate which in turn affect the 
behavior on the surface. In fact, the process of epitaxial growth
is a nonequilibrium phenomenon in which surface, interfacial and elastic
energies along with surface diffusion and alloy segregation play
active roles. During the nonequilibrium growth of alloy films, elastic 
stress may be relieved through nucleation of dislocations and other 
defects. Inhomogeneities can also arise on account of two potential 
instabilities, from which the generation of dislocations can be avoided 
and one can obtain high quality thin film materials.
These are the morphological instability and alloy segregation instability.

Morphological instability \cite{asaro,grinfeld,srolovitz} can be understood
through two features: A solid film adsorbate on a solid substrate creates a
uniaxial stress, and  the driving force for surface diffusion of matter is
proportional to the local strain energy density. If a small corrugation
occurs on the surface of a uniaxially strained solid, local strain energy
concentration is created at its valleys. As a result diffusion of matter 
occurs along the surface from valleys to peaks of the corrugation. Valleys
then deepen due to loss of mass which leads to a further increase of
local strain energy driving the instability. Surface energy plays a
stabilizing role here: as corrugations deepen, surface area increases 
which has an associated energy cost inhibiting the growth of instability. 

The alloy segregation instability is driven by alloy thermodynamics. 
\cite{cahn1} It is typically modelled by a Ginzburg-Landau type free 
energy (the order parameter $\phi$ is related to the binary alloy 
concentration) in which the quadratic term has a coefficient proportional 
to $(T-T_c)$. This term changes sign for $T < T_c$ and leads to an 
instability which drives a homogeneous mixture to segregate. This 
mechanism is called spinodal decomposition and is applicable for not 
too asymmetric mixtures within the classical spinodal. For asymmetric 
mixtures between the spinodal and coexistence curves, the mixture is 
metastable and requires large amplitude fluctuations to segregate via 
a nucleation mechanism.

In this paper we consider strained film growth without nucleation, and
assume the growing film to be a coherent, dislocation-free binary or
pseudo-binary alloy. Stability of such films have been considered
previously both theoretically \cite{tersoff1,tersoff2,tersoff3,guyer1,%
guyer2,guyer4,guyer5,spencer1,spencer2,spencer3,spencer4,spencer5,glas,%
leonard1,leonard2,leonard3,leonard4,leonard5} and experimentally. The 
systems explored experimentally include SiGe \cite{xie1,jesson1,schelling,%
perovic,lagally1,tromp,walther} and InGaAs. \cite{snyder,okada,peiro,%
gonzalez,guyer3,gendry,grandjean,chokshi} Other interesting experimental 
studies include: single component films (Ge on a Si substrate),
\cite{legoues} GaAs homoepitaxy,\cite{tiedje} self-organised GaAs islands,
\cite{xie2} etc.; but these are not as relevant here. Many of these 
experimental papers show that stressed growing films can develop non-planar 
morphologies without dislocations or nucleation, for both symmetric binary 
alloy films (50-50 mixture) as well as asymmetric films. The onset of 
instability is measured through a kinetic critical thickness $h_c$ 
(film thickness at which surface roughness first appears), and the 
characteristic wavelength of the instability has also been determined
by experiments.

Early theoretical work focused on explaining nonplanar surface
morphologies through the competition between surface energy and 
elastic strain energy arising from lattice mismatch $\epsilon = 
(\bar{a}^f - a^s) / a^s$, where $\bar{a}^f$ is the  average bulk 
film lattice parameter and $a^s$ is the substrate lattice spacing.
The important effects due to composition dependence of the film 
lattice parameter $a^f$, represented by the solute expansion 
coefficient \cite{cahn1,cahn2} $\eta = (\partial a^f /\partial 
\phi)/a^f$ where $\phi$ is the local concentration field, were 
added in a kinetic treatment of the growing film. \cite{guyer1,%
guyer2,guyer4,guyer5,spencer1,spencer2,spencer3,leonard1,leonard3} 
Intricate dependence on $\epsilon$ and $\eta$ 
leads to sensitivity of the morphological development to the sign 
of the misfit $\epsilon$, \textit{i.e.} compressive and tensile 
regions of the growing film behave differently. Another important 
factor, the composition dependence of the elastic moduli, can also 
impact on the stability of the film and its sensitivity to the sign 
of $\epsilon$.\cite{leonard1,leonard3} Experimental observations 
can provide a sensitive test of these theories in connection with 
the $\epsilon$-asymmetry in the stability diagram of the growing 
film, the coupling between morphological instability and alloy 
segregation, as well as the behaviors of critical thickness.

An additional factor is the differing atomic mobilities of the two 
(or more) species on the growing film surface, and has been 
investigated in recent theoretical work.\cite{tersoff1,tersoff2,%
spencer1,spencer2} This consideration is based on recent 
experimental results of atomic mobilities on surface. E.g., the
activation energy of step mobility for Ge on Ge(001) surface was 
measured to be smaller than that of Si on Si(001), \cite{chey}
corresponding to much larger atomic mobility for Ge. However, 
during the MBE growth for semiconductors (e.g., Si-Ge system or 
III-V alloys), dimerized atoms are quickly formed on (001) top 
surface when new materials are deposited, leading to a typical
phenomenon of surface reconstruction.\cite{note-dimer} These dimers, 
which may be comprised of unlike alloy components or like atoms, 
diffuse over the surface with diffusion processes and properties 
more complicated than and different from that of monomers.  
\cite{swartz} For example, in Ge-on-Si submonolayer system
the surface pattern with Si-Ge mixed dimers rather than Ge-Ge dimers
can be observed.\cite{qin} A mixed Si-Ge dimer can diffuse as a unit, 
or transform into a pure dimer between like atoms (Si-Si) due to 
atomic exchange with underlying substrate atom and can also return
back to the mixed form due to re-exchange process.\cite{swartz,%
qin} Therefore, different alloy components deposited on surface
may or may not diffuse independently. In the following studies 
we consider an effective surface mobility (or effective surface
diffusivity $D_s$) to describe the processes of surface diffusion
and decomposition in continuum approximation.

In this paper our starting point is the work of L\'{e}onard and 
Desai. It is reviewed in Ref.~\onlinecite{leonard6} and the basic 
model on which some refinements are made is described in 
Ref.~\onlinecite{leonard3}. Here we further develop the model
to a more general case, and more importantly, we consider two 
key features in the epitaxial growth systems: The first one is
the coupling between top surface and underlying bulk 
of the growing film, which leads to a new dynamic equation for
the evolution of surface compositional variable ($\phi_s$); the 
second one is the composition dependence of all the three elastic
constants, including the Young's modulus $E$, shear modulus $\mu$ 
and Poisson ratio $\nu$ ($=E/2\mu-1$). In Ref.~\onlinecite{leonard3}, 
the shear modulus was assumed to be constant and the surface-bulk 
coupling was neglected in dynamical equations. These new 
considerations can lead to stability results different from 
that of the previous theoretical work for strained alloy films,
\cite{guyer1,guyer2,guyer4,spencer1,spencer2,spencer3,leonard1,%
leonard3} and more complicated and richer properties of the system 
can be obtained. Here we are interested in the growing films with
deposition rate $v\neq 0$, while the results for isothermal 
annealing alloy films without growth ($v=0$) are presented elsewhere.
\cite{huang} The details of our model for the growth of thin 
alloy films are described in Sec. \ref{sec:II}. In Sec. \ref{sec:III} 
we present the solution of mechanical equilibrium for the system
(More general results for mechanical equilibrium solution are given 
in Appendix \ref{appendA}). In Sec. \ref{sec:IV} we use the results
of Sec. \ref{sec:III} to derive self-contained dynamic equations for 
morphological and compositional perturbations, which we then use to 
perform a linear stability analysis; this leads to the characteristic 
equation for perturbation growth rate. The formulae 
of elastic free energy in Fourier space, which are needed in
the derivation of Sec. \ref{sec:IV}, are given in Appendix
\ref{appendB}. The results of stability analysis for both
the composition independent and composition dependent elastic moduli,
including the stability diagrams, are presented in Sec. \ref{sec:V}.
The behavior of kinetic critical thickness is presented in 
Sec. \ref{sec:VI}. Finally we conclude with a summary of salient
features of our results in Sec. \ref{sec:VII}.

\section{\label{sec:II}Model}

Let us consider a strained alloy system composed of a semi-infinite
substrate occupying the region $z<\zeta(x,y)$ and a A$_{1-X}$B$_X$ 
binary or pseudo-binary alloy film in the region $\zeta(x,y)<z<h(x,y,t)$. 
Here $\zeta(x,y)$ refers to the vertical position at which the 
film-substrate interface is located, and $h(x,y,t)$ is the surface 
height variable of the growing film. Usually, $\zeta(x,y)$ is assumed 
to be constant, or for simplicity $0$, corresponding to a planar 
film-substrate interface fixed at $z=0$.\cite{nonzero-zeta}
The local composition within the film can be denoted by a continuous 
variable $\phi({\bf r},t)$ which is proportional to the difference 
in the local concentrations of two alloy constituents. Its average 
value $\bar{\phi}$ is equal to $2X-1$, with $X$ the alloy composition. 
Here we focus on the symmetric mixture film, i.e., $X=1/2$ 
alloy for which $\bar{\phi}=0$, while in the substrate $z\le \zeta$
we have $\phi=0$ since the compositional fluctuation is usually
assumed to be absent there.

For this system of epitaxial growth, the film is assumed to be
elastically isotropic and coherent with the substrate, 
without any misfit dislocations or other defects. On the top 
surface of the growing film, the evaporation and re-condensation 
are negligible if the deposition of material occurs under 
ultra-high vacuum condition, like the growth through MBE, and 
then the system evolution should correspond to conserved dynamics.
We also assume that there is no interdiffusion between film 
and substrate, and the diffusion and compositional relaxation 
in the bulk film can be neglected, since the bulk mobility of
film components is miniscule compared to that on the surface in 
typical epitaxial growth. The neglect of the bulk mobility 
implies that the composition profile within the film reflects 
the time-history of the growing film and is dependent on the 
deposition rate $v$. As well the layers within the film are buried 
metastable layers which are frozen on account of negligible mobility 
in the bulk. Therefore, the bulk concentration field $\phi_b$ at 
some time $t$ is equal to the surface field $\phi_s$ of an earlier 
time, i.e., $\phi_b(x,y,z,t) = \phi_s(x,y,t-t_0)$ for $z<h$, 
where $t_0\sim(\bar{h}-z)/v$ with $\bar{h}=vt$ the average surface 
height, and then we have
\begin{equation}
\phi_b(x,y,z,t) \sim \phi_s(x,y,t=z/v).
\label{eq-phi_b}
\end{equation}

Consequently, the conserved dynamics of morphological and 
compositional evolution is dominated by the surface diffusion 
and surface decomposition processes, with two time-dependent 
essential dynamical variables: the surface morphology $h(x,y,t)$
and the concentration field at the surface $\phi(x,y,z=h(x,y,t),t)
\equiv \phi_s(x,y,t)$. The evolution equation describing the time 
dependence of $h$ is \cite{leonard3,leonard5}
\begin{equation}
\frac{\partial h}{\partial t} = \Gamma_h \sqrt{g} \nabla_s^2
\frac{\delta {\cal F}}{\delta h} +v,
\label{eq-h}
\end{equation}
where $v$ represents the deposition rate or growth velocity, 
$\nabla_s^2$ is the surface Laplacian, $g=1+|\nabla h|^2$ 
denotes the determinant of the surface metric, and $\Gamma_h=
D_s N_s/k_B T N_v^2$ is the kinetic coefficient \cite{mullins} 
which depends on the surface diffusivity $D_s$, the Boltzmann 
constant $k_B$, growth temperature $T$, as well as the number 
density of atoms per unit surface area $N_s$ and that per unit
volume $N_v$. For the concentration field $\phi$, the conserved
dynamics leads to
\begin{equation}
\frac{\partial \phi}{\partial t} = \Gamma_{\phi} \nabla^2
\frac{\delta {\cal F}}{\delta \phi}-\Lambda \phi \Delta(z-h),
\label{eq-phi}
\end{equation}
where $\Lambda$ is proportional to film deposition rate $v$, 
corresponding to the tendency of surface composition to go towards
the homogeneous phase (of $X=1/2$ and $\bar{\phi}=0$ in this
paper) driven by the uniform deposition flux of material,
\cite{leonard5,atzmon} and $\Delta(z-h)$ is zero everywhere
except on the top surface $z=h$, where it is $1$. 
Here the mobility $\Gamma_{\phi}$ is approximately
zero in bulk, while at surface it is $\Gamma_{\phi}=%
\Gamma_h\delta^{-1}$ with $\delta$ the effective diffusion 
thickness of surface layer. The total free energy functional 
$\cal{F}$ in Eqs. (\ref{eq-h}) and (\ref{eq-phi}) makes the
evolution of variables $h$ and $\phi$ coupled with each other
and consists of three contributions:
\begin{equation}
\cal{F}=\cal{F}_{\rm s} + \cal{F}_{\rm GL} +
\cal{F}_{\rm el},
\label{eq-F}
\end{equation}
with the surface energy $\cal{F}_{\rm s}$, the Ginzburg-Landau
free energy $\cal{F}_{\rm GL}$, and the elastic energy 
$\cal{F}_{\rm el}$. The addition of deposition noise and thermal 
noise terms in 
the dynamical equations for $h$ and $\phi$ have been considered
earlier. \cite{leonard5} At the level of linear stability 
analysis, neglecting noise does not alter the results.
In Eq. (\ref{eq-phi}) there are two ways to evaluate the surface 
composition fluctuations. The first one \cite{leonard3} is only 
considering the surface state and surface free energy, that is, 
evaluating the free energy ${\cal F}$ at surface $z=h$ and then 
calculating the functional differentiation directly with respect 
to surface composition field $\phi_s(x,y,t)$ to obtain its evolution 
behavior. In this paper, we also take into account the coupling 
between surface and underlying bulk of the growing film by (i) 
identifying the free energy ${\cal F}$ as the total energy of the 
system, that is, from $z=-\infty$ to $z=h$, and not just the one 
evaluated at surface $z=h$, (ii) then calculating $\nabla^2%
\delta {\cal F}/\delta \phi$ as a whole, and finally (iii) 
evaluating its value as well as the other terms in Eq. (\ref{eq-phi}) 
at surface $z=h$ to obtain the dynamics of surface field $\phi_s$. 
This is analogous in spirit to the procedure in the previous study 
of surface critical phenomena for spin fluctuations.\cite{kumar}

The surface energy contribution $\cal{F}_{\rm s}$ to the free energy 
(\ref{eq-F}) plays a stabilizing role and can be represented by a
drumhead model without pinning term:
\begin{equation}
{\cal F}_{\rm s}[h]=\gamma \int d^2r \sqrt{g},
\label{eq-F_s}
\end{equation}
where $\gamma$ is the surface tension at the top surface assumed to 
be isotropic and composition independent.\cite{note-gamma} The alloy 
thermodynamics is enforced through the Ginzburg-Landau functional
\begin{equation}
{\cal F}_{\rm GL}[\phi,h]=\int_{\zeta}^{h} d^3r
\left [-{\frac{r'}{2}}\phi^2 + {\frac{u}{4}}\phi^4
+ {\frac{c}{2}}|\nabla\phi|^2 \right ],
\label{eq-F_GL}
\end{equation}
where the coefficients can be expressed as $r'=k_B(T_c-T)N_v$ 
and $c=k_BT_cN_v\lambda_0^2 /2$, with $T_c$ the critical temperature 
of the binary alloy and $\lambda_0$ the interaction distance which
is of the order of lattice spacing.\cite{cahn3} In the absence of
elastic effect, when $T>T_c$ the equilibrium state of bulk alloy
is homogeneous with $\phi=0$, while for $T<T_c$ two phases 
$\phi=\pm \sqrt{r'/u}$ coexist in equilibrium. The gradient 
energy term in Eq. (\ref{eq-F_GL}) is important for stability 
analysis. It penalizes the sharp compositional changes and without 
this term a nonphysical divergence for short wavelength mode will 
occur.\cite{guyer2,guyer5,spencer1} The coefficient $c$ of this
gradient energy term is related to the interfacial tension arising
at the interface between coexisting phases.

The elastic free energy functional $\cal{F}_{\rm el}$ is important
in any stress-driven system such as a growing film. Linear 
elasticity theory provides its expression as
\begin{equation}
{\cal F}_{\rm el}[\phi,{\bf u},h]=\frac{1}{2}\int_{-\infty}^{h}
d^3r S_{ijkl}\sigma_{ij}\sigma_{kl},
\label{eq-F_el}
\end{equation}
where $\sigma_{ij}$ is the stress tensor and $S_{ijkl}$ is
the elastic compliance tensor with subscripts $i$, $j$, $k$, 
or $l=x$, $y$, $z$. For the isotropic system assumed here, 
we have $S_{ijkl}=\delta_{ik}\delta_{jl}(1+\nu)/E
-\delta_{ij}\delta_{kl}\nu/E$, with Young's modulus $E$ 
and Poisson ratio $\nu$. Generally, the elastic constants of 
the film depend on the local composition. In the first order 
approximation they are
\begin{eqnarray}
E^f&=&E^f_0(1 + E_1^* \phi), \nonumber\\
\mu^f&=&\mu^f_0(1 + \mu_1^* \phi),
\label{eq-Eu}
\end{eqnarray}
and $\nu^f=E^f/2\mu^f-1$ (with $\mu$ the shear modulus), and 
the composition-independent results can be obtained by setting 
$E_1^* = \mu_1^* =0$. In this paper, superscript $f$ or $s$ 
refers to the film or substrate respectively, and in most of 
our calculations below (except Appendix \ref{appendA}) we neglect 
the difference between the average elastic constants of the film 
($E^f_0$, $\mu^f_0$, and $\nu^f_0=E^f_0/2\mu^f_0-1$) and of the 
substrate ($E^s$, $\mu^s$, and $\nu^s$). This corresponds to the 
material systems with substrate and film having similar elastic 
constants.

Using the model developed here and the equations (\ref{eq-h})--%
(\ref{eq-Eu}), we can explore the stability of the growing symmetric 
alloy films in the parameter space of lattice mismatch $\epsilon$, 
solute expansion coefficient $\eta$, deposition rate $v$, growth
temperature $T$, and the composition dependent elastic moduli 
parameters $E_1^*$ and $\mu_1^*$.

\section{\label{sec:III}Solution of Mechanical Equilibrium}

Thermodynamic equilibrium consists of mechanical, thermal and chemical
equilibria. In the nonequilibrium system of growing alloy films that
we are considering, the mechanical equilibrium occurs on a very fast
time scale and we can assume that it occurs instantaneously. Thus
by solving the equations of mechanical equilibrium, we can obtain the
elastic displacement field ${\bf u}$ in terms of the other two fields
$\phi$ and $h$. This is the goal of this section. Knowing this solution,
its substitution in the elastic free energy functional $\cal{F}_{\rm el}$
renders it dependent only on $\phi$ and $h$. The resulting effective
elastic free energy functional is then used in the dynamic equations
(\ref{eq-h}) and (\ref{eq-phi}) for further analysis.

Inside both the film and substrate, the mechanical equilibrium yields
\begin{equation}
\partial_j \sigma_{ij}=0,
\label{eq-equi}
\end{equation}
where the linear elastic stress tensor $\sigma_{ij}$ obeys the Hooke's 
law, and for the isotropic system studied here it is given by
\begin{equation}
\sigma^f_{ij} = 2\mu^f \left [\frac{\nu^f}{1-2\nu^f}u^f_{ll}\delta_{ij}
+u^f_{ij} - \frac{1+\nu^f}{1-2\nu^f}(\epsilon +\eta\phi)\delta_{ij} 
\right ]
\label{eq-stress_f}
\end{equation}
in the film, subjected to both the misfit strain $\epsilon$ and
compositional strain $\eta\phi$, as well as 
\begin{equation}
\sigma^s_{ij} = 2\mu^s \left [\frac{\nu^s}{1-2\nu^s}u^s_{ll}\delta_{ij}
+u^s_{ij} \right ]
\label{eq-stress_s}
\end{equation}
in the substrate. Here $u_{ij}=(\partial_i u_j + \partial_j u_i)/2$ 
is the linear elastic strain tensor, and the reference state we use 
is the unconstrained state of substrate bulk lattice.

Considering the negligible external pressure on top surface of the
film and the continuity of displacement and stress at the coherent 
film-substrate interface, we have the corresponding boundary 
conditions: 
\begin{equation}
\sigma_{ij}^f n^f_j =0 \qquad {\rm at} \quad z=h(x,y,t), 
\label{eq-bound1}
\end{equation}
with $n^f_j$ the unit vector normal to the surface, and
\begin{equation}
\sigma_{ij}^f n_j =\sigma_{ij}^s n_j
\qquad {\rm and} \qquad u_i^f = u_i^s  
\qquad {\rm at} \quad z=\zeta(x,y),
\label{eq-bound2}
\end{equation}
where $n_j$ is the unit normal to the film-substrate interface
and oriented toward the film. Moreover, in the substrate the 
displacement and strain are expected to vanish far from the 
interface, \textit{i.e.}
\begin{equation}
u_i^s\rightarrow 0 \qquad {\rm and} \qquad u_{ij}^s\rightarrow 0
\qquad {\rm for} \quad z\rightarrow-\infty.
\label{eq-bound3}
\end{equation}

To solve the above mechanical equilibrium equation, all the variables
are expanded in Fourier series with a general form
\begin{equation}
\xi =\bar{\xi} + \sum\limits_{\bf q}\hat{\xi}({\bf q},z,t)
e^{i(q_x x+q_y y)},
\label{eq-expan}
\end{equation}
where $\xi$ could represent different variables such as $\phi$, $h$,
$u_i$, and $\zeta$, and $\hat{\xi}({\bf q},z,t)$ denotes the small 
perturbation around the basic state solution $\bar{\xi}$ (for height
variable $h$ or interface variable $\zeta$, the perturbation should
be expressed as $\hat{h}({\bf q},t)$ or $\hat{\zeta}({\bf q})$,
respectively). In the basic state, we have a uniform growing film 
with homogeneous composition $\bar{\phi}=0$ and a planar front moving 
at constant rate $v$, corresponding to film thickness $\bar{h}=vt$.
Due to the coherency of the film, the in-plane film lattice constant
is equal to $a^s$ and then $\bar{u}_x^f=\bar{u}_y^f=0$. The Poisson
relaxation in the $z$ direction results in $\bar{u}_z^f=\bar{u}^f_{zz}z$,
with $\bar{u}^f_{zz}=\bar{u}=\epsilon (1+\nu^f_0)/(1-\nu^f_0)$. The film
is stressed in the lateral directions, leading to $\bar{\sigma}_{xx}^f
=\bar{\sigma}_{yy}^f=\bar{\sigma}=-2\mu^f_0\bar{u}$, and all the other
strains and stresses are zero. In the substrate, we have $\bar{u}_i^s
=0$ and $\bar{u}_{ij}^s=\bar{\sigma}_{ij}^s=0$ for $i$, $j=x$, $y$, $z$, 
corresponding to an unstrained basic state.

The expansion forms of both the mechanical equilibrium (\ref{eq-equi}) 
and boundary conditions (\ref{eq-bound1})--(\ref{eq-bound3}) can be 
obtained using Eq. (\ref{eq-expan}). Generally, due to composition
dependence of film elastic constants, as depicted in Eq. (\ref{eq-Eu}),
from Eqs. (\ref{eq-equi}) and (\ref{eq-stress_f}) one can obtain the 
extra terms proportional to coefficients $E_1^*$ and $\mu_1^*$ in the 
mechanical equilibrium equation of the film, which is different from 
the composition independent case and more complicated (since the 
elastic constants also vary with position ${\bf r}$ according to 
Eq. (\ref{eq-Eu})). Here we only consider the solution of mechanical
equilibrium to zeroth order of elastic moduli $E^f$ and $\mu^f$,
which is applicable when $E_1^*$ and $\mu_1^*$ are small enough,
or $E^f$ and $\mu^f$ are assumed to vary very slowly with space
and the equation is considered to first order.
Consequently, we obtain the mechanical equilibrium equation with 
a linear form as used before \cite{leonard3}, but with
different linearized boundary conditions due to the film-substrate
interface roughness $\hat{\zeta}$. At the top surface of the film,
\textit{i.e.} $z=\bar{h}$, from Eq. (\ref{eq-bound1}) we have
\begin{eqnarray}
\hat{\sigma}^f_{xz} &=& iq_x \bar{\sigma} \hat{h}, \nonumber\\
\hat{\sigma}^f_{yz} &=& iq_y \bar{\sigma} \hat{h}, \nonumber\\
\hat{\sigma}^f_{zz} &=& 0,
\label{eq-bound1-q}
\end{eqnarray}
as before, while at interface $z=\bar{\zeta}=0$, we have different
conditions due to Eq. (\ref{eq-bound2}):
\begin{eqnarray}
&& -iq_x \bar{\sigma} \hat{\zeta} + \hat{\sigma}^f_{xz} = 
\hat{\sigma}^s_{xz}, \nonumber\\
&& -iq_y \bar{\sigma} \hat{\zeta} + \hat{\sigma}^f_{yz} = 
\hat{\sigma}^s_{yz}, \nonumber\\
&& \hat{\sigma}^f_{zz} = \hat{\sigma}^s_{zz}, 
\label{eq-bound21-q}
\end{eqnarray}
and
\begin{eqnarray}
&& \hat{u}^f_i = \hat{u}^s_i \qquad {\rm for} \quad i=x,y,
\nonumber\\
&& \bar{u} \hat{\zeta} + \hat{u}^f_z = \hat{u}^s_z.
\label{eq-bound22-q}
\end{eqnarray}
Finally, for $z\rightarrow -\infty$ condition (\ref{eq-bound3}) 
leads to
\begin{equation}
\hat{u}_i^s\rightarrow 0 \qquad {\rm and} \qquad 
\partial_z \hat{u}_i^s\rightarrow 0. 
\label{eq-bound3-q}
\end{equation}
The corresponding results of the solution for non-planar interface 
$\zeta \neq 0$ and differing film-substrate elastic constants are 
detailed in Appendix \ref{appendA}. In the following studies for
elastic free energy and dynamical equations, we assume a planar
film-substrate interface,\cite{nonzero-zeta} \textit{i.e.} 
$\zeta=0$, as well as the equal average elastic constants between
film and substrate, that is, $E^f_0 = E^s = E_0$, $\mu^f_0 =
\mu^s = \mu_0$, and $\nu^f_0 = \nu^s = \nu_0$. Thus, substituting
these conditions in the results of Appendix \ref{appendA}, one
can obtain the solution
\begin{equation}
\hat{u}^f_i = \left [
\begin{array}{c}
\alpha_x \\ \alpha_y \\ \alpha_z 
\end{array} \right ] 
e ^{qz} - \left [
\begin{array}{c}
iq_x/q \\ iq_y/q \\ 1
\end{array} \right ] 
Cz e ^{qz} + \left ( \frac{1+\nu_0}{1-\nu_0} \right ) \eta \left [
\begin{array}{c}
iq_x \hat{W} \\ iq_y \hat{W} \\ \partial_z \hat{W}
\end{array} \right ], 
\label{eq-u^f}
\end{equation}
for the film, where $q^2=q_x^2 +q_y^2$, and the coefficients $\alpha_i$
and $C$ are given by
\begin{eqnarray}
&iq_x \alpha_x + iq_y \alpha_y = e^{-q\bar{h}} & \left [ 
(2(1-\nu_0)-q\bar{h}) \left ( q\bar{u}\hat{h} + 
\left ( \frac{1+\nu_0}{1-\nu_0} \right ) \eta q
\left ( \partial_z \hat{W} \right )_{z=\bar{h}} \right ) 
\right. \nonumber\\
&&\left. +(-1+2\nu_0+q\bar{h})\left ( \frac{1+\nu_0}{1-\nu_0} \right ) 
\eta \left ( \partial^2_z \hat{W} - \hat{\phi} \right )_{z=\bar{h}} 
\right ], 
\label{eq-axy}
\end{eqnarray}
\begin{eqnarray}
&q\alpha_z = e^{-q\bar{h}} & \left [ 
(1-2\nu_0+q\bar{h}) \left ( q\bar{u}\hat{h} + 
\left ( \frac{1+\nu_0}{1-\nu_0} \right ) \eta q
\left ( \partial_z \hat{W} \right )_{z=\bar{h}} \right ) 
\right. \nonumber\\
&&\left. -(2(1-\nu_0)+q\bar{h})\left ( \frac{1+\nu_0}{1-\nu_0} \right ) 
\eta \left ( \partial^2_z \hat{W} - \hat{\phi} \right )_{z=\bar{h}} 
\right ], 
\label{eq-az}
\end{eqnarray}
\begin{equation}
C = e^{-q\bar{h}} \left [ q\bar{u}\hat{h} + 
\left ( \frac{1+\nu_0}{1-\nu_0} \right ) \eta 
\left ( q \partial_z \hat{W} - \partial^2_z \hat{W} + \hat{\phi}
\right )_{z=\bar{h}} \right ], 
\label{eq-C}
\end{equation}
with $q_y \alpha_x = q_x \alpha_y$. Here $\hat{W}$ is introduced
through the particular solution of the mechanical equilibrium
equation, as shown in the last term of Eq. (\ref{eq-u^f}), and
is defined through
\begin{equation}
(\partial_z^2-q^2)\hat{W}=\hat{\phi},
\label{eq-W}
\end{equation}
or the relation $\nabla^2 W =\phi$ in real space.\cite{leonard3} 
Consequently, the results for strain and stress tensors can be 
obtained through e.g., Eq. (\ref{eq-u_lz}).

\section{\label{sec:IV}Dynamic Equations and Perturbation Growth Rate}

To evaluate the dynamic equations (\ref{eq-h}) and (\ref{eq-phi})
and then determine the evolution behaviors of surface profile,
we should first calculate the total free energy ${\cal F}$. 
The first two contributions, surface and Ginzburg-Landau free
energy, are given in Eqs. (\ref{eq-F_s}) and (\ref{eq-F_GL})
explicitly, while the film elastic energy can be obtained from 
Eqs. (\ref{eq-F_el}) and (\ref{eq-stress_f}):
\begin{eqnarray}
& {\cal F}^f_{\rm el} =& \int_0^{h} d^3r 
\left [ \frac{1}{2}\lambda^f {u^f_{ll}}^2 + \mu^f {u^f_{ij}}^2
\right. \nonumber\\
&& \left. +\frac{E^f}{1-2\nu^f} \left (\frac{3}{2} \epsilon^2
-\epsilon u^f_{ll} +3\epsilon \eta \phi 
+\frac{3}{2} \eta^2 \phi^2 -\eta \phi u^f_{ll} \right )\right ],
\label{eq-F_el^f}
\end{eqnarray}
with the Lam\'{e} coefficient $\lambda^f=2\mu^f \nu^f/(1-2\nu^f)$.
In Eq. (\ref{eq-F_el^f}), the first two terms represent the general
form of the isotropic elastic energy density for single-component
system, and by setting $\epsilon=0$ and renormalizing the 
Ginzburg-Landau free energy, which absorbs the $\phi^2$ term,
we can obtain the bulk elastic energy functional with the effect
of compositional strain (\textit{i.e.} coupling term $\eta \phi 
u^f_{ll}$) that has been widely used. \cite{leonard4,onuki} 
Considering the composition dependence of both Young's and shear 
moduli, we can write Eq. (\ref{eq-F_el^f}) to first order of 
$E_1^*$ and $\mu_1^*$ as
\begin{eqnarray}
& {\cal F}^f_{\rm el} =& \int_0^{h} d^3r 
\left \{ \frac{1}{2}\lambda_0 {u^f_{ll}}^2 + \mu_0 {u^f_{ij}}^2
+ \mu_0 \mu_1^* \phi {u^f_{ij}}^2 \right. \nonumber\\
&& +\frac{E_0}{1-2\nu_0} \left [ \frac{3}{2} \epsilon^2
-\epsilon u^f_{ll} + 3 \left ( \epsilon \eta + 
\frac{3E_1^*-2(1+\nu_0)\mu_1^*}{2(1-2\nu_0)} \epsilon^2 \right )
\phi  \right. \nonumber\\
&& + 3\left ( \frac{1}{2} \eta^2 + \frac{3E_1^*-2(1+\nu_0)\mu_1^*}
{1-2\nu_0} \epsilon \eta \right ) \phi^2 
-\left ( \eta + \frac{3E_1^*-2(1+\nu_0)\mu_1^*}{1-2\nu_0}
\epsilon \right ) \phi u^f_{ll} \nonumber\\
&& +\frac{1}{2(1-2\nu_0)} 
\left ( E_1^* - \frac{1+2\nu_0^2}{1+\nu_0}\mu_1^* \right )
\phi {u^f_{ll}}^2 -\frac{3E_1^*-2(1+\nu_0)\mu_1^*}{1-2\nu_0}
\eta \phi^2 u^f_{ll} \nonumber\\
&&\left. \left. + \frac{3}{2} \left( 
\frac{3E_1^*-2(1+\nu_0)\mu_1^*}{1-2\nu_0} \right )
\eta^2 \phi^3 \right ] \right \}
\label{eq-F_el-Eu}
\end{eqnarray}
with $\lambda_0=2\mu_0 \nu_0/(1-2\nu_0)$. The corresponding 
Fourier expansion to second order of perturbation is given in 
Appendix \ref{appendB}.

The strain tensor $u^f_{ij}$ in elastic energy expression 
(\ref{eq-F_el-Eu}) can be determined from the solution of
mechanical equilibrium shown in Eqs. (\ref{eq-u^f})--(\ref{eq-C})
or Appendix \ref{appendA}, and is in fact composed of two
parts:
\begin{equation}
u_{ij}^f = u_{ij}^{hom} + u_{ij}^{par}.
\label{eq-u_ij}
\end{equation}
The first part is the homogeneous solution and can be calculated 
through the first two terms of the displacement solution 
(\ref{eq-u^f}). The second part corresponds to the particular 
solution and is given by 
\begin{equation}
u_{ij}^{par} = \frac{1+\nu_0}{1-\nu_0}\eta
\partial_i \partial_j W.
\label{eq-u^par}
\end{equation}
Note that the homogeneous part $u_{ij}^{hom}$ is in fact not the 
function of system variable $\phi$ or $W$, and the coefficients in 
this part, $\alpha_i$ and $C$ in Eqs. (\ref{eq-axy})--(\ref{eq-C}),
depend on $(\partial_z \hat{W})_{z=\bar{h}}$, 
$(\partial^2_z \hat{W})_{z=\bar{h}}$, and $\hat{\phi}|_{z=\bar{h}}$
which are surface values (of fixed $z$ for certain time $t$)
evaluated at $z=\bar{h}$ due to the boundary condition at top 
surface. Therefore, when performing the functional differentiation
of the free energy with respect to system variable $\phi$, only the
particular solution (\ref{eq-u^par}) of strain tensor needs to
be considered. \cite{note-dev}

The dynamical equation for compositional perturbation at the surface 
can be determined by substituting energy formulae (\ref{eq-F_s}), 
(\ref{eq-F_GL}) and (\ref{eq-F_el-Eu}) into Eq. (\ref{eq-phi}). 
When expanded to first order, it is
\begin{equation}
\partial\hat{\phi}_s/\partial t = -\Gamma_{\phi} q^2
(-r' + c q^2) \hat{\phi}_s + \Gamma_{\phi} \left. \left ( 
\nabla^2 \frac{\delta {\cal F}^f_{\rm el}}{\delta \phi} 
\right )_q \right |_s - \Lambda \hat{\phi}_s,
\label{eq-phi_tq}
\end{equation}
with the elastic energy term $(\nabla^2 \delta {\cal F}^f_{\rm el}
/ \delta \phi )_q |_s$ obtained from the Fourier expansion using
Eq. (\ref{eq-F_el-Eu}):
\begin{eqnarray}
\left. \left ( \nabla^2\frac{\delta {\cal F}^f_{\rm el}}
{\delta \phi} \right )_q \right |_s = -q^2 && \left \{ 
\left [ \frac{E_0(3-\nu_0)}{(1-\nu_0)^2} \eta^2 +
\frac{E_0(2(5-7\nu_0)E_1^*-(7-10\nu_0)(1+\nu_0)\mu_1^*)}
{(1-\nu_0)^2(1-2\nu_0)}\epsilon\eta \right ] \hat{\phi}_s
\right. \nonumber\\
&& \left. - \left [ \frac{E_0}{1-\nu_0}\eta
+\frac{E_0(2E_1^*-(1+2\nu_0)\mu_1^*)}{(1-2\nu_0)(1-\nu_0)}
\epsilon \right ] \hat{u}^f_{ll}|_s +\frac{E_0}{1-\nu_0}
\mu_1^* \epsilon \hat{u}^f_{zz}|_s \right \},
\label{eq-deltaF-q}
\end{eqnarray}
where we have used the continuity condition of $\nabla^2\phi$
at surface: $(\nabla^2\phi)_s = \nabla^2_s \phi_s$, which
implies that $[(\partial_z^2 \phi)_s]_q \approx 0$ since
we ignore terms of order $\hat{h}^2 \hat{\phi}_s$.

For the evolution of surface height variable $h$, we can
obtain the following linearized dynamical equation by
combining Eq. (\ref{eq-h}) and the free energy functionals
(\ref{eq-F_s}), (\ref{eq-F_GL}) and (\ref{eq-F_el-Eu}):
\begin{equation}
\partial\hat{h}/\partial t = -\Gamma_h q^2 \left (
\gamma q^2 \hat{h} + \hat{\cal E}^f|_s \right ),
\label{eq-h_tq}
\end{equation}
where $\hat{\cal E}^f|_s$ is the first order elastic free energy
density evaluated at the surface, with the expression given in
Eq. (\ref{eq-F1}) of Appendix \ref{appendB}.

Since what we are interested in is the stability property of
this growing alloy system with respect to small perturbations,
we focus on the film behavior of the early evolution time
regime, where the morphological and compositional perturbations 
at the surface are assumed to evolve exponentially:
\begin{equation}
\hat{h}=\hat{h}_0(q) e^{\Omega_h t} \qquad {\rm and} 
\qquad \hat{\phi}_s = \hat{\phi}_0(q) e^{\Omega_{\phi} t},
\label{eq-Omega}
\end{equation}
with $\Omega_h$ and $\Omega_{\phi}$ the growth rates of surface 
perturbations for $h$ and $\phi$, respectively. Due to the 
coupling of dynamical equations (\ref{eq-h}) and (\ref{eq-phi})
in general case, the joint stability or instability, corresponding
to $\Omega_h=\Omega_{\phi}=\Omega$, is obtained. Considering the 
expression (\ref{eq-phi_b}) for bulk composition field, we have 
\cite{leonard3} $\hat{\phi}=\hat{\phi}_0 \exp(\Omega z/v)$, and 
then $\hat{W}=v^2 \hat{\phi} /(\Omega^2-q^2v^2)$ from Eq. 
(\ref{eq-W}). Consequently, by substituting this approximate 
form of $\hat{W}$ into the solution for film displacement vector 
$\hat{u}^f_i$ (\ref{eq-u^f})--(\ref{eq-C}), we can calculate formula 
(\ref{eq-deltaF-q}) as well as the linear elastic energy density
$\hat{\cal E}^f|_s$ in term of two surface fields $\hat{h}$ and
$\hat{\phi}_s$, as shown in Appendix \ref{appendB}. 

For further analysis, it is convenient to introduce a characteristic 
length scale representing the typical size of interfaces between surface 
domains:
\begin{equation}
l_0 = \left ( \frac{|r|}{c}\right ) ^{-1/2},
\label{eq-l0}
\end{equation}
as well as a diffusive time scale
\begin{equation}
\tau_0 = \left (\Gamma_{\phi}\frac{r^2}{c}\right )^{-1},
\label{eq-tau0}
\end{equation}
where $r=r'-2E_0\eta^2/(1-\nu_0)=k_B(T_c^{\rm eff}-T)N_v$
is a renormalized coefficient with the effective critical
temperature
\begin{equation}
T_c^{\rm eff}=T_c - \frac{2E_0}{1-\nu_0}\frac{\eta^2}{k_BN_v}.
\label{eq-Tceff}
\end{equation}
Note that $T_c^{\rm eff}$ used here is just the bulk coherent
spinodal temperature derived by Cahn \cite{cahn1} for binary 
stressed alloys. Using the following transformations
\begin{eqnarray}
k &=& ql_0, \nonumber\\
\tau &=& t/\tau_0, \nonumber\\
V &=& v \tau_0 / l_0 = \Lambda \tau_0, \nonumber\\
\sigma &=& \Omega \tau_0, \nonumber\\
\gamma^* &=& \gamma l_0 / c, \nonumber\\
\epsilon^* &=& \left [\frac{2E_0}{|r|}\left (\frac{1+\nu_0}
{1-\nu_0}\right ) \right ]^{1/2} \epsilon, \nonumber\\
\eta^* &=& \left [\frac{2E_0}{|r|}\left (\frac{1+\nu_0}
{1-\nu_0}\right ) \right ]^{1/2} \eta,
\label{eq-scale}
\end{eqnarray}
where we have set $\Gamma_h / \Gamma_{\phi} = \delta =l_0$, 
as well as
\begin{eqnarray}
\hat{h}^* &=& h/l_0, \nonumber\\
\hat{\phi}_s^* &=& \hat{\phi}_s,
\label{eq-scale2}
\end{eqnarray}
we reduce the dynamical equations (\ref{eq-h_tq}) and (\ref{eq-phi_tq}) 
for surface perturbations to dimensionless forms:
\begin{equation}
\partial\hat{h}^*/\partial \tau=({\epsilon^*}^2k^3-\gamma^*k^4)
\hat{h}^* - \frac{k^2}{1+\nu_0}\left [ \epsilon^*\eta^* 
\left ( \frac{\sigma-kV\nu_0}{\sigma+kV} \right ) + 
\frac{2E_1^*-(1+\nu_0)\mu_1^*}{2(1-\nu_0)}{\epsilon^*}^2 \right ]
\hat{\phi}_s^*,
\label{eq-h1}
\end{equation}
and
\begin{eqnarray}
\partial\hat{\phi}_s^*/\partial \tau&=&\frac{k^3}{1-\nu_0}\left [
(1-2\nu_0)\epsilon^*\eta^* + (2E_1^*-(1+\nu_0)\mu_1^*){\epsilon^*}^2
\right ]\hat{h}^* \nonumber\\
&+& \left [ -k^2 \left ( k^2\pm 1+\frac{8E_1^*-5(1+\nu_0)\mu_1^*}
{2(1-\nu_0^2)}\epsilon^*\eta^* \right ) -V \right. \nonumber\\
&+& \left. \frac{1}{1-\nu_0} \left ( (1-2\nu_0){\eta^*}^2
+(2E_1^*-(1+\nu_0)\mu_1^*)\epsilon^*\eta^* \right )
\frac{k^3 V}{\sigma+kV} \right ] \hat{\phi}_s^*,
\label{eq-phi1}
\end{eqnarray}
where for sign ``$\pm$'', the top sign applies when the system is
above the effective critical temperature $T_c^{\rm eff}$ and the
bottom sign is taken if $T<T_c^{\rm eff}$.

Due to the rescaling (\ref{eq-scale}), the early time behavior of
perturbations (\ref{eq-Omega}) is transformed to $\hat{h}^*=
\hat{h}^*_0\exp(\sigma_h\tau)$ and $\hat{\phi}_s^*=\hat{\phi}^*_0
\exp(\sigma_{\phi}\tau)$ with the nondimensional perturbation 
growth rates $\sigma_h$ and $\sigma_{\phi}$. Generally, the dynamical
equations (\ref{eq-h1}) and (\ref{eq-phi1}) for $\hat{h}^*$ and
$\hat{\phi}_s^*$ couple with each other, resulting in a cubic
characteristic equation for perturbation growth rate $\sigma_h=
\sigma_{\phi}=\sigma$:
\begin{eqnarray}
&&(\sigma+\gamma^*k^4-{\epsilon^*}^2 k^3)\left [ \sigma
+k^2\left (k^2\pm1+\frac{8E_1^*-5(1+\nu_0)\mu_1^*}{2(1-\nu_0^2)}
\epsilon^*\eta^* \right ) \right . \nonumber\\
&&\left. -\frac{1}{1-\nu_0} \left ( (1-2\nu_0){\eta^*}^2
+(2E_1^*-(1+\nu_0)\mu_1^*)\epsilon^*\eta^* \right )
\frac{k^3 V}{\sigma+kV} +V \right ] \nonumber\\
&&+\frac{k^5}{1-\nu_0^2}\left [ (1-2\nu_0)\epsilon^*\eta^* 
+ (2E_1^*-(1+\nu_0)\mu_1^*){\epsilon^*}^2 \right ] \nonumber\\
&&\times\left [ \epsilon^*\eta^* \left ( \frac{\sigma-kV\nu_0}
{\sigma+kV} \right ) + \frac{2E_1^*-(1+\nu_0)\mu_1^*}
{2(1-\nu_0)} {\epsilon^*}^2 \right ] =0.
\label{eq-sigma}
\end{eqnarray}
When the real part of $\sigma$ is larger than zero, the initial
perturbations will grow with time, leading to the instability
of the growing film; otherwise the surface perturbations will
decay in time, making the system stable. Note that this 
characteristic equation (\ref{eq-sigma}) is asymmetric with respect 
to the sign of rescaled misfit $\epsilon^*$, leading to the 
asymmetry of stability properties between compressive and tensile 
films as specified below, but it is symmetric for $(\epsilon^*,
\eta^*) \rightarrow (-\epsilon^*,-\eta^*)$, similar to the previous 
results. \cite{guyer2,spencer1} Thus, in the following analysis
we keep $\eta^* \geq 0$ and consider different signs of
$\epsilon^*$.

Setting $E_1^*=\mu_1^*=0$ in above deductions we can obtain the 
corresponding results for system with composition independent 
elastic constants. The corresponding equation for $\hat{h}^*$ 
is the same as that for $\hat{h}^*$ in Ref. \onlinecite{leonard3}, 
but the equations for ${\phi}_s^*$ are quite different due to the 
fact that here we consider the total energy of the film and the 
surface-bulk coupling, while in Ref. \onlinecite{leonard3} only 
the surface energy is used (see also Appendix \ref{appendB}). 
Consequently, the characteristic equations for $\sigma$ are also 
different, and our Eq. (\ref{eq-sigma}) reduces to a cubic,
which is also different from that of the previous related work
carried out by Guyer and Voorhees \cite{guyer1,guyer2,guyer4,guyer5}
and Spencer \textit{et al.} \cite{spencer1,spencer2,spencer3}
due to different models and approximations. Compared with Eq. 
(\ref{eq-sigma}), this cubic equation shows the symmetry 
with respect to the sign of $\epsilon^*$ for systems with
composition independent elastic moduli.

\section{\label{sec:V}Stability diagrams}

\begin{figure}
\resizebox{0.6\textwidth}{!}{
\includegraphics{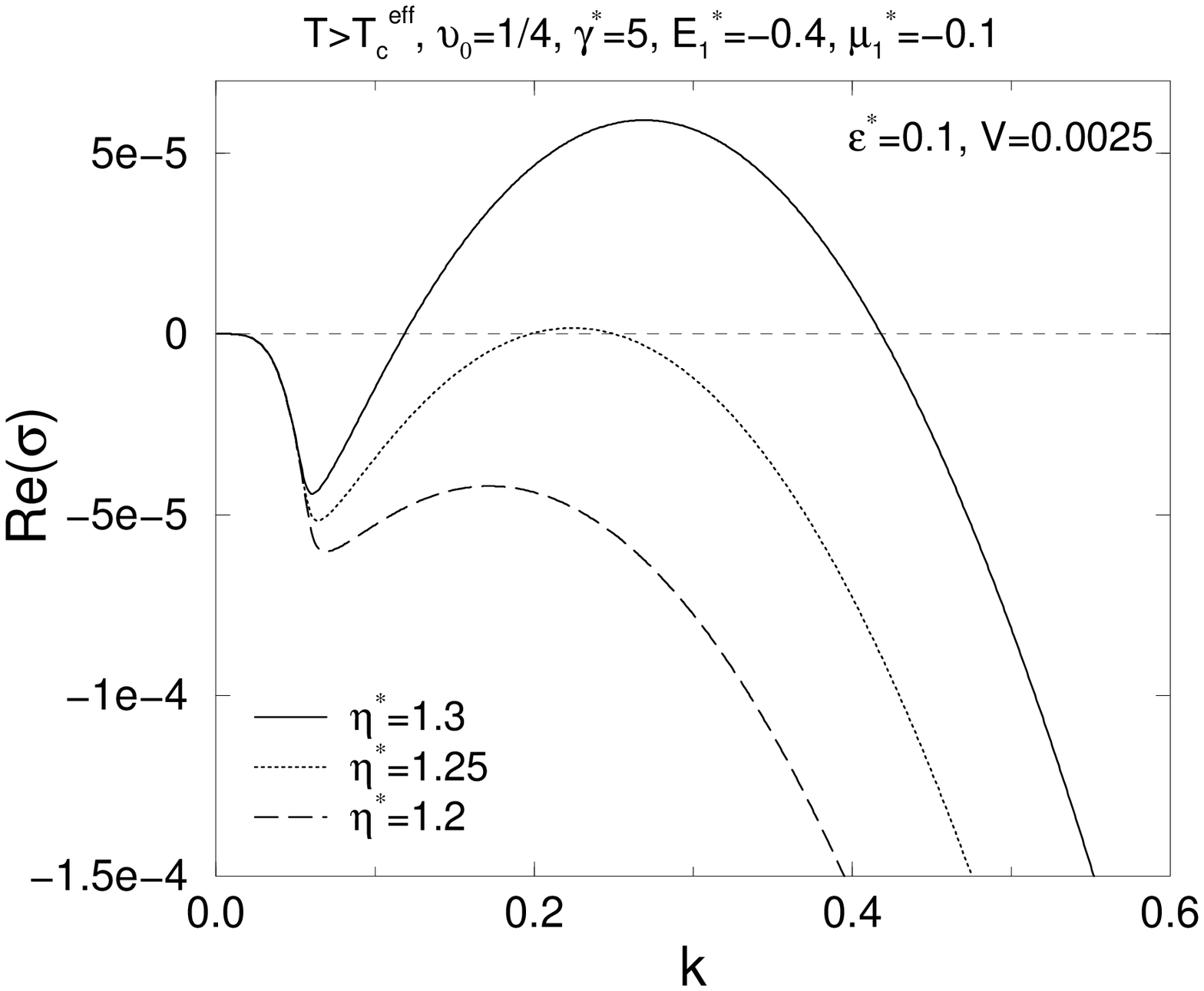}} 
\caption{\label{fig-sig}Dispersion relation for real part of
perturbation growth rate $\sigma$ versus wavenumber $k$.
}
\end{figure}

The typical dispersion relations of Re$(\sigma)$, which is 
the real part of joint perturbation growth rate $\sigma$, 
as a function of rescaled wavenumber $k$ are depicted in Fig.
\ref{fig-sig} for systems with composition dependent elastic
moduli and $T>T_c^{\rm eff}$. The value of Re$(\sigma)$ in Fig. 
\ref{fig-sig} for each $k$ is in fact defined as the largest 
real part of the three solutions of characteristic equation 
(\ref{eq-sigma}), since that will be the dominant growth in 
case of instability. Note that for very large $k$ values we 
always have Re$(\sigma)\rightarrow -\infty$, even when $\eta^*$ 
is large, corresponding to short wavelength stabilization. This 
is different from the previous work \cite{guyer2,guyer5,spencer1} 
where nonphysical divergence for $k\rightarrow \infty$ and large 
$\eta^*$ occurs due to the absence of gradient term in free energy 
functional. If Re$(\sigma)$ is smaller than zero for all $k$, e.g., 
the dashed curve in Fig. \ref{fig-sig}, the system is stable; 
otherwise, we can find a band of wavenumbers for which the surface 
profiles are unstable to fluctuations. In the following we 
investigate the stability properties and the effects of misfit 
strain, compositional stress, deposition rate, as well as the 
composition dependence of elastic constants more systematically, 
by determining the maximum of Re$(\sigma)$ with respect to
all disturbance modes $k$ for different material parameters.

\subsection{\label{sec:V-A}Analytic results for special cases}

We first consider some simple cases where the misfit strain
or solute stress is absent and then the analytic results can 
be available. The simplest case corresponds to $\epsilon^*=
\eta^*=0$, \textit{i.e.} a lattice-matched film with  constituents 
having the same size, for which dynamical equations (\ref{eq-h1}) 
and (\ref{eq-phi1}) decouple, resulting in different perturbation 
growth rates for $\hat{h}^*$ and ${\phi}_s^*$, as found in previous 
work:
\cite{leonard3} 
\begin{eqnarray}
\sigma_h &=& -\gamma^* k^4, \label{eq-sig-h1}\\
\sigma_{\phi} &=& -k^4 \mp k^2 -V.
\label{eq-sig-phi1}
\end{eqnarray}
Formula (\ref{eq-sig-h1}) implies the stability of film morphology
since $\sigma_h$ is always negative. Eq. (\ref{eq-sig-phi1}) yields
the stable compositional profile for $T>T_c^{\rm eff}$ (top sign
``$-$''), whereas for $T<T_c^{\rm eff}$ the competition between 
the deposition of homogeneous materials and surface decomposition
leads to a critical deposition rate $V_c=1/4$. When $V>V_c$ the
film is stable with respect to compositional fluctuations, otherwise
phase separation occurs. 

When misfit strain exists ($\epsilon^*\neq 0$) but solute stress
is absent ($\eta^*=0$), we have different results for systems with
composition independent and composition dependent elastic moduli.
When $E_1^*=\mu_1^*=0$, the height variable and composition field
still decouple, and we can recover the result of Asaro and Tiller
\cite{asaro} and Grinfeld \cite{grinfeld} for morphological 
instability, which is
\begin{equation}
\sigma_h = {\epsilon^*}^2 k^3 - \gamma^* k^4.
\label{eq-sig-h2}
\end{equation}
The stability of compositional perturbation is also governed by 
Eq. (\ref{eq-sig-phi1}), in accord with the previous result.
\cite{leonard3} However, when the elastic constants are dependent 
on film composition, $E_1^*\neq 0$ and $\mu_1^*\neq 0$ make the 
dynamical equations for $\hat{h}^*$ and ${\phi}_s^*$ coupled, and 
then the joint perturbation growth rate $\sigma_h = \sigma_{\phi} 
=\sigma$ obeys a quadratic equation
\begin{equation}
\sigma^2 + a_1 \sigma + a_0 =0,
\label{eq-quad}
\end{equation}
with coefficients
\begin{eqnarray}
a_1 &=& \gamma^* k^4 - {\epsilon^*}^2 k^3 + k^2(k^2\pm 1)+V, 
\nonumber\\
a_0 &=& (\gamma^* k^4 - {\epsilon^*}^2 k^3) [k^2(k^2\pm 1)+V]
+ \frac{[2E_1^*-(1+\nu_0)\mu_1^*]^2}{2(1+\nu_0)(1-\nu_0)^2}
{\epsilon^*}^4 k^5.
\label{eq-a10-1}
\end{eqnarray}
Note that the solution form for quadratic (\ref{eq-quad}) is
$\sigma=\left (-a_1\pm\sqrt{a_1^2-4a_0}\right )/2$, and we have 
maximum Re$(\sigma)>0$ due to $a_1>0$ and $a_0<0$ obtained from 
(\ref{eq-a10-1}) when $k\ll 1$, for both $T$ above and
below $T_c^{\rm eff}$. Therefore, for the lattice mismatched
film without solute stress, the morphological profile and 
compositional mode couple with each other if we consider the 
composition dependence of elastic moduli, and then the misfit 
strain can not only cause the morphological instability, but 
also the phase separation in spite of uniform deposition, 
which is different from the previous stability results. 
\cite{spencer3,leonard3}

So far in this subsection, we have considered different cases
all of which have $\eta^* =0$. From Eq. (\ref{eq-Tceff}), we
note that for these cases $T_c^{\rm eff}$ is same as $T_c$.

\begin{figure}
\resizebox{0.6\textwidth}{!}{
\includegraphics{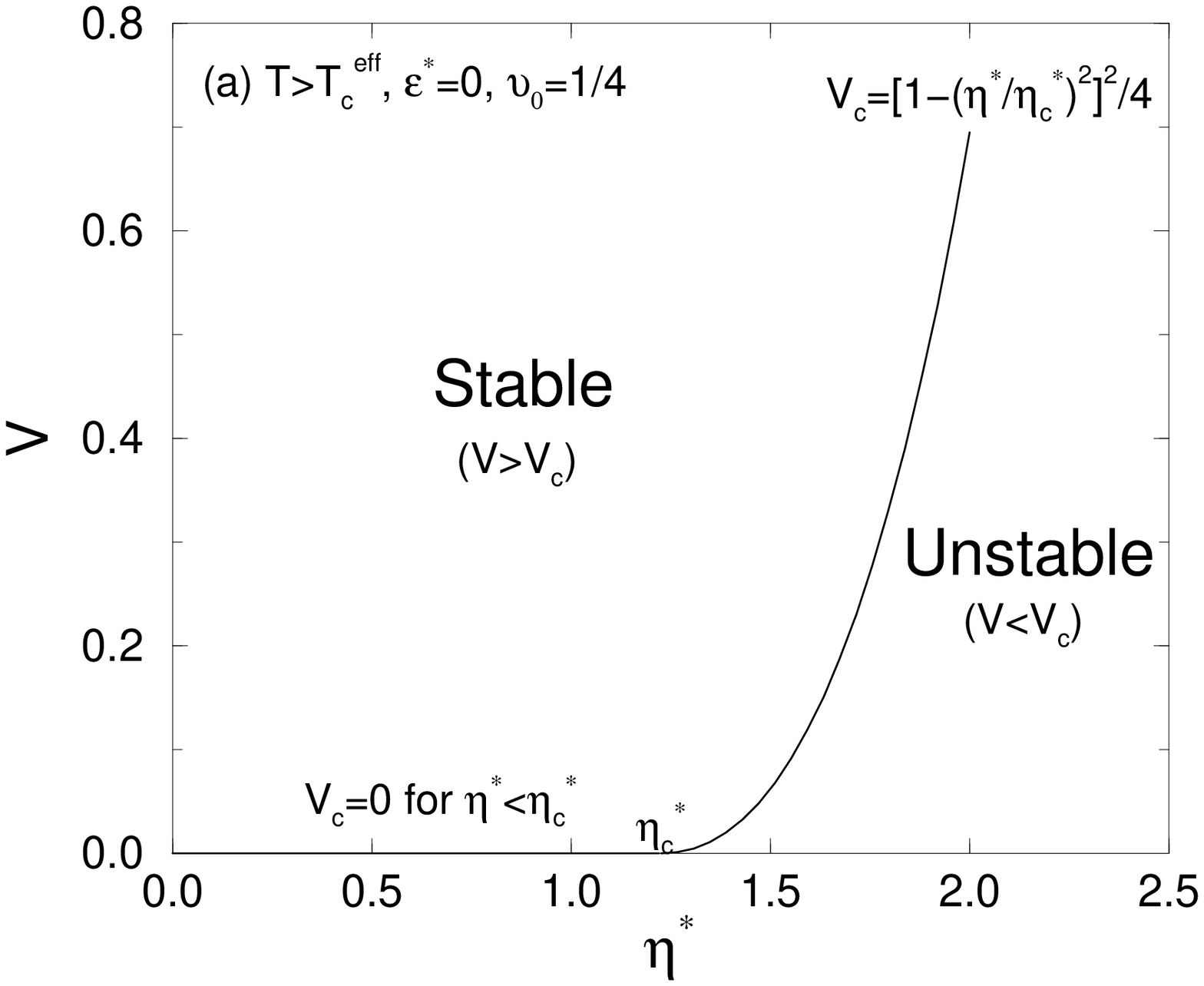}} \\
\resizebox{0.6\textwidth}{!}{
\includegraphics{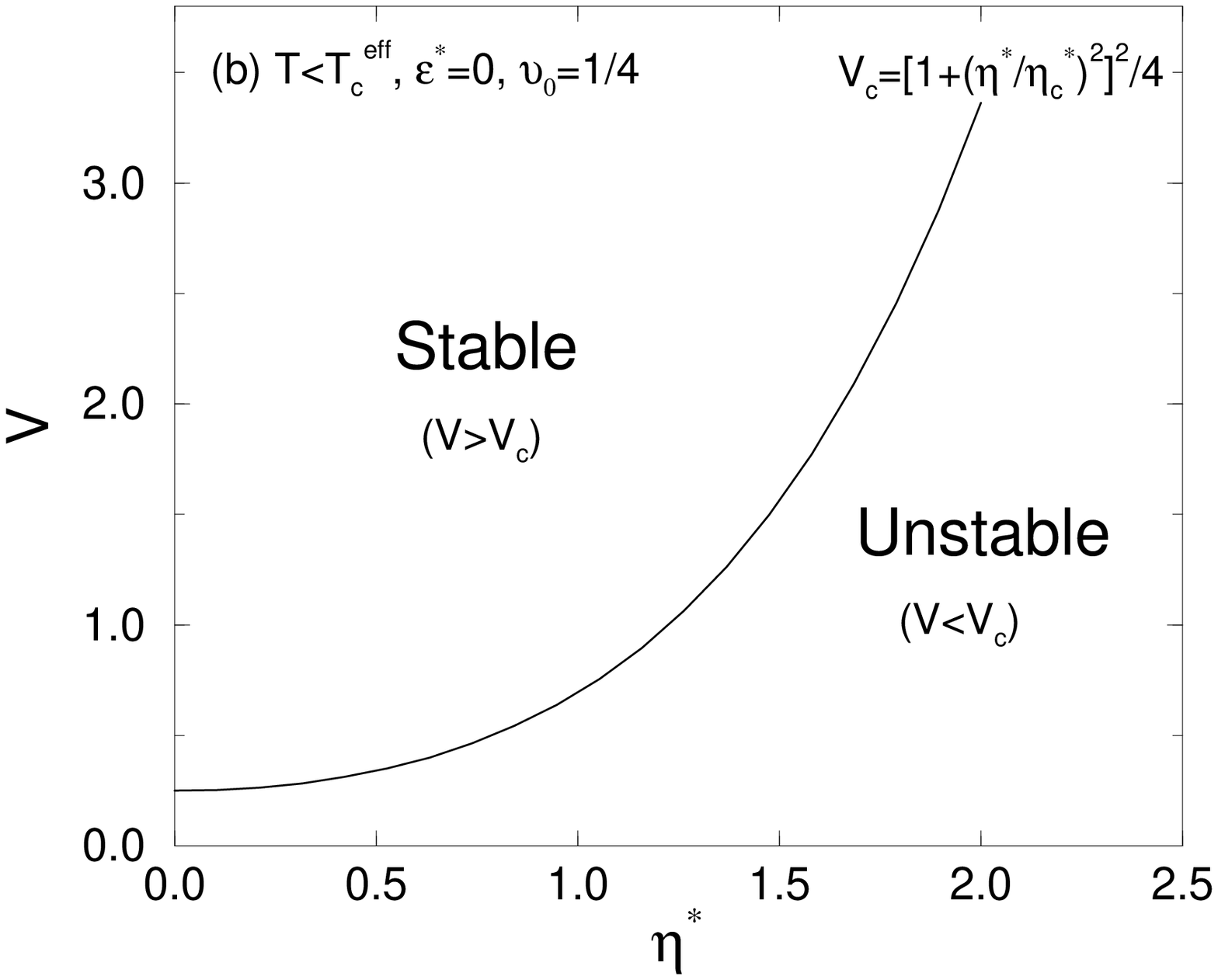}} 
\caption{\label{fig-eps0}Stability diagrams of rescaled 
deposition rate $V$ versus $\eta^*$, for special case of no 
misfit $\epsilon^*=0$, $\nu_0=1/4$, as well as growth
temperature (a) $T>T_c^{\rm eff}$ and (b) $T<T_c^{\rm eff}$.
These diagrams are calculated from the analytic results given
in Eqs. (\ref{eq-st-eps0+}) and (\ref{eq-st-eps0-}).
}
\end{figure}

The case for no misfit strain ($\epsilon^*=0$) but nonzero solute
stress ($\eta^*\neq 0$) can determine the role of compositional 
strain on film stability (without the interplay with misfit). The 
same results of perturbation rate can be obtained from dynamical 
equations for composition dependent (Eqs. (\ref{eq-h1}) and 
(\ref{eq-phi1})) and composition independent elastic moduli. 
That is, the morphological and compositional modes decouple, 
with $\sigma_h$ given by Eq. (\ref{eq-sig-h1}), resulting in 
stable morphology of this lattice-matched film. The property of 
compositional perturbation is more complicated, with $\sigma_{\phi}$ 
governed by a quadratic similar to (\ref{eq-quad}):
$$\sigma_{\phi}^2 + a_1 \sigma_{\phi} + a_0 =0,$$
where the coefficients are
\begin{eqnarray}
a_1 &=& k^2(k^2\pm 1) +(k+1)V, \nonumber\\
a_0 &=& kV \left [ k^2(k^2\pm 1) - k^2 \left (\frac{1-2\nu_0}
{1-\nu_0} \right ) {\eta^*}^2 +V \right ]. 
\label{eq-a10-2}
\end{eqnarray}
The stability conditions for compositional profile can then be
calculated analytically through the solutions of this quadratic 
equation for $\sigma_{\phi}$. For growth temperature $T$ higher 
than bulk coherent spinodal $T_c^{\rm eff}$, which corresponds to 
the case of most experiments, phase separation can occur only
when $\eta^*$ is larger than a critical value $\eta_c^*$, while
for $\eta^* < \eta_c^*$ the lattice-matched alloy film is stable
against decomposition. Different condition applies to low
temperature $T<T_c^{\rm eff}$, for which there is no critical
value like $\eta_c^*$ for the appearance of alloy decomposition,
and larger deposition rate $V$ can suppress the instability.
In more detail, the \textit{stability} condition for $T>
T_c^{\rm eff}$ is given as
\begin{eqnarray}
&& {\eta^*}^2 < {\eta_c^*}^2 =
\frac{1-\nu_0}{1-2\nu_0} \qquad {\rm for \ any} \ V \nonumber\\
&& {\rm or} \nonumber\\
&& {\eta^*}^2 > {\eta_c^*}^2 \quad {\rm and} \quad
V > \frac{1}{4} \left [1-\left (\eta^* / \eta_c^* \right )^2 
\right ]^2.
\label{eq-st-eps0+}
\end{eqnarray}
For $T<T_c^{\rm eff}$ the compositional profile is stabilized
with the condition of high deposition rate
\begin{equation}
V > \frac{1}{4} \left [1+\left (\eta^* / \eta_c^* \right )^2 
\right ]^2,
\label{eq-st-eps0-}
\end{equation}
which is fulfilled for any $\eta^*$, with the value of $\eta_c^*$ 
given in Eq. (\ref{eq-st-eps0+}). The corresponding stability 
diagrams calculated from Eqs. (\ref{eq-st-eps0+}) and 
(\ref{eq-st-eps0-}) are plotted in Fig. \ref{fig-eps0}, with 
$\nu_0=1/4$ for which ${\eta_c^*}^2=1.5$. In both Fig. 
\ref{fig-eps0} (a) and Fig. \ref{fig-eps0} (b), the compositional
profile is stable in the region of high deposition rate $V>V_c$,
with $V_c$ generally increasing with the value of $\eta^*$ except
for a special region of $T>T_c^{\rm eff}$. This special region
corresponds to $\eta^* < \eta_c^*$ as given in Eq. (\ref{eq-st-eps0+}),
and in this range of $\eta^*$ we have $V_c=0$, that is, the system
is always stable.

Therefore, our results for this instability driven by compositional  
stress, in the absence of misfit, are different from those of 
earlier works, \cite{guyer5,spencer1} where the compositional 
instability is expected only below an effective temperature which 
increases with $\eta^2$, but is higher than the coherent spinodal 
$T_c^{\rm eff}$ used here and independent of deposition rate $V$.
Our calculations exhibit more complicated behavior, that is, the 
solute stress enhances the compositional modulation only when 
exceeding a critical value ($\eta_c^*$) for $T>T_c^{\rm eff}$, and 
this stress driven instability can be suppressed by large 
deposition rate, as illustrated in Fig. \ref{fig-eps0} as well
as Eqs. (\ref{eq-st-eps0+}) and (\ref{eq-st-eps0-}). Note that 
both the increase form of $V_c$ and the value of $\eta_c^*$ 
in Fig. \ref{fig-eps0} are different from that of Ref. 
\onlinecite{leonard3}, due to different models and characteristic 
equations for $\sigma_{\phi}$.

\subsection{\label{sec:V-B}Stability diagrams in general case}

Generally, both misfit strain and solute stress are nonzero in
strained alloy film growth, corresponding to $\epsilon^* \neq 0$,
$\eta^* \neq 0$, and the coupling of dynamical equations for
surface height and compositional variables. Thus, the growth of
morphological disturbance will induce the decomposition and phase
separation of alloy film, and vice versa, resulting in the 
simultaneous occurrence of stability or instability for morphological
and compositional profiles. The joint perturbation growth rate
is then determined by the cubic characteristic equation 
(\ref{eq-sigma}), which can be rewritten as
\begin{equation}
\sigma^3 + a_2\sigma^2 +a_1\sigma +a_0 =0,
\label{eq-cubic}
\end{equation}
with the three coefficients:
\begin{eqnarray}
a_2 &=& \gamma^* k^4 - {\epsilon^*}^2 k^3 + 
k^2\left ( k^2\pm1+\beta_0\epsilon^*\eta^* \right )
+(k+1)V, \nonumber\\
a_1 &=& (\gamma^* k^4 - {\epsilon^*}^2 k^3 + kV)
\left [k^2(k^2\pm 1+\beta_0 \epsilon^*\eta^*) +V \right ] \nonumber\\
&+& kV (\gamma^* k^4 - {\epsilon^*}^2 k^3) -
\frac{k^3 V}{1-\nu_0} \left [ (1-2\nu_0){\eta^*}^2 
+\alpha_0\epsilon^*\eta^* \right ] \nonumber\\
&+& \frac{k^5}{2(1+\nu_0)(1-\nu_0)^2}\left [ (1-2\nu_0)
\epsilon^*\eta^* + \alpha_0 {\epsilon^*}^2
\right ] \left [ 2(1-\nu_0)\epsilon^*\eta^* + 
\alpha_0 {\epsilon^*}^2 \right ], \nonumber\\
a_0 &=& (\gamma^* k^4 - {\epsilon^*}^2 k^3) \left \{
kV \left [ k^2(k^2\pm 1+\beta_0 \epsilon^*\eta^*) +V \right ]
-\frac{k^3 V}{1-\nu_0} \left [ (1-2\nu_0){\eta^*}^2 
+\alpha_0\epsilon^*\eta^* \right ] \right \} \nonumber\\
&+& \frac{k^6 V}{2(1+\nu_0)(1-\nu_0)^2}\left [ (1-2\nu_0)
\epsilon^*\eta^* + \alpha_0 {\epsilon^*}^2 \right ] 
\left [ -2\nu_0(1-\nu_0)\epsilon^*\eta^* + 
\alpha_0 {\epsilon^*}^2 \right ].
\label{eq-a210-1}
\end{eqnarray}
Here the parameters $\alpha_0$ and $\beta_0$ are defined as
\begin{eqnarray}
\alpha_0 &=& 2E_1^*-(1+\nu_0)\mu_1^*, \nonumber\\
\beta_0 &=& \frac{8E_1^*-5(1+\nu_0)\mu_1^*}{2(1-\nu_0^2)}.
\label{eq-alpha-beta}
\end{eqnarray}

We can determine the stability properties with respect to 
different parameters such as $\epsilon^*$, $\eta^*$, $V$, and 
$T$ from the largest real-part root of above characteristic 
equations, which can be obtained from the dispersion relations 
similar to those shown in Fig. \ref{fig-sig}. In this and the 
following sections two sets of material parameters are used, 
corresponding to two typical types of stability properties.
In the first set, we choose $\nu_0=1/4$, $\gamma^*=5$, $\eta^*$
around $0.6$, as well as $E_1^*=-0.4$ and $\mu_1^*=-0.1$ if
considering the composition dependent elastic moduli, which
qualitatively represents the alloy films of SiGe type.
The second set corresponds to the systems of pseudo-binary
III-V alloys such as InGaAs, with the chosen parameters
$\nu_0=1/3$, $\gamma^*=3.5$, $\eta^*$ around $1.1$, and
$E_1^*=-0.25$, $\mu_1^*=-0.5$ if applied. These parameter 
values have been arrived at by using the rescaling formulae
(\ref{eq-scale}) as well as different types of material 
parameters for group IV and III-V components presented in Ref. 
\onlinecite{guyer1} (e.g., the elastic moduli and the atomic
size difference for alloy components that determines $\eta$). 
The parameters such as the elastic constants and lattice parameters 
are estimated from the average of the quantities for pure components 
(e.g., Si and Ge or InAs and GaAs), while $E_1^*$ and $\mu_1^*$ are 
adjustable but should be qualitatively consistent with the variation 
trend of $E$ and $\nu$ between pure alloy components.

\begin{figure}
\resizebox{\textwidth}{!}{
\includegraphics{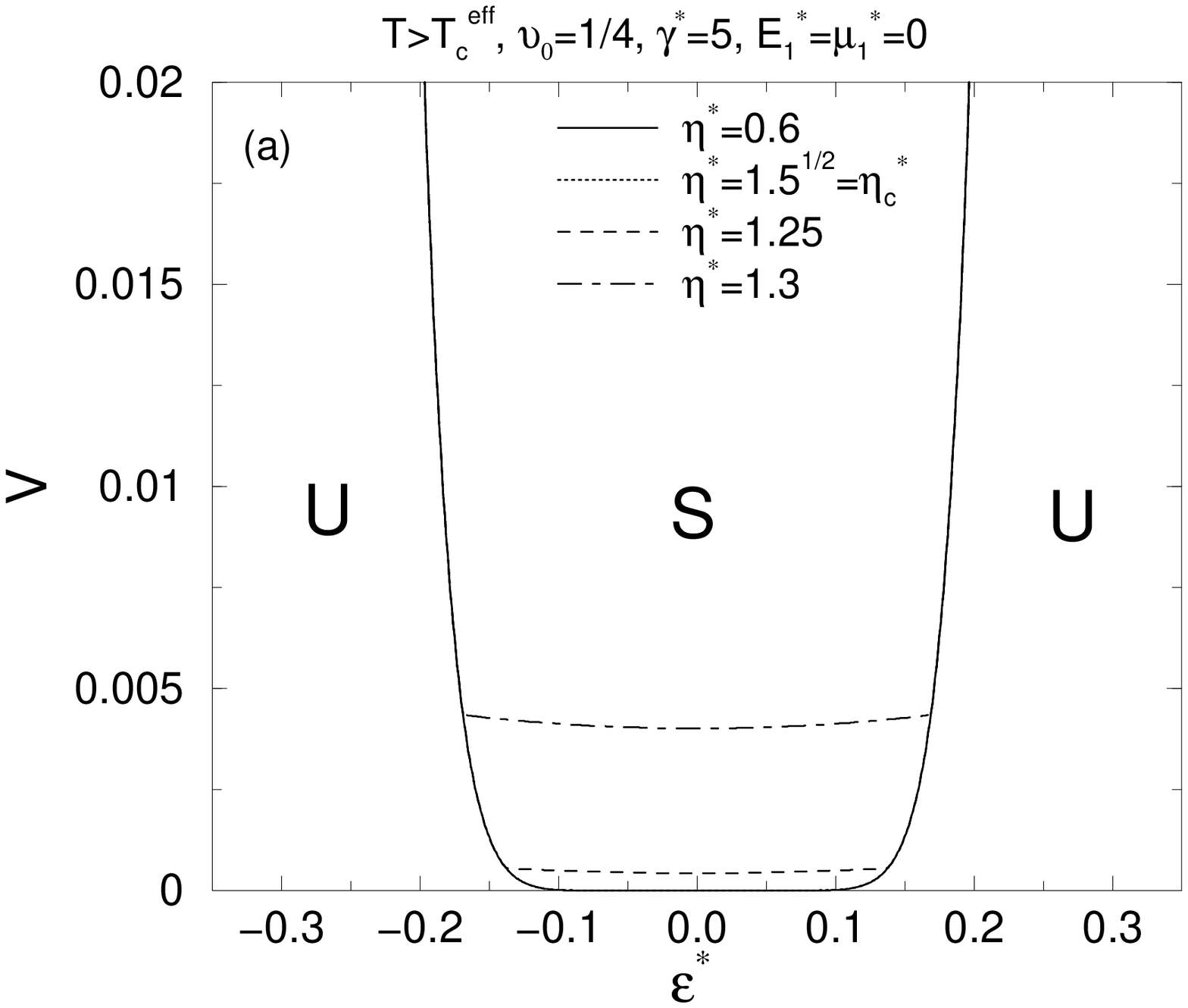}
\hskip 1.5cm
\includegraphics{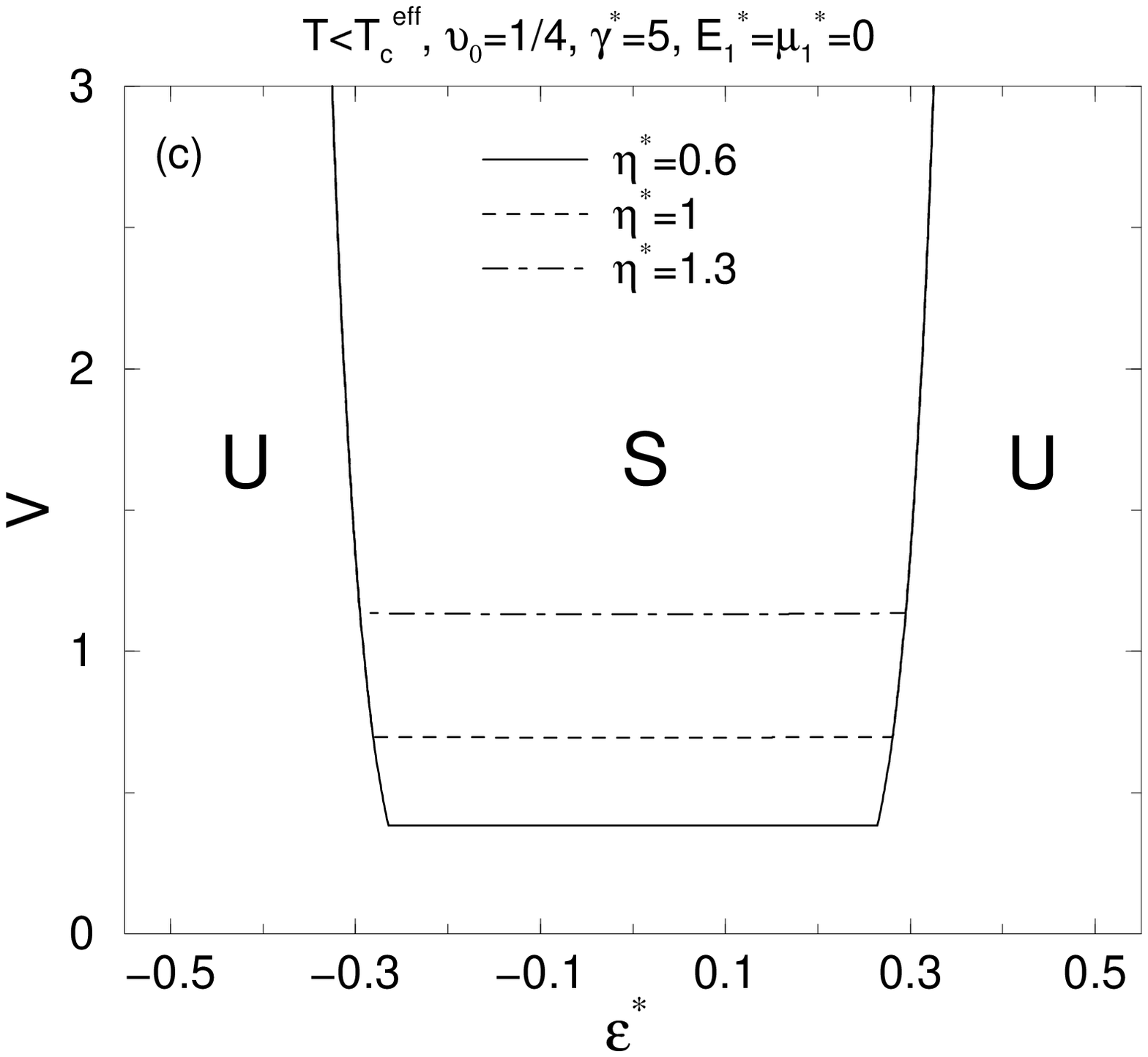}} \\
\resizebox{\textwidth}{!}{
\includegraphics{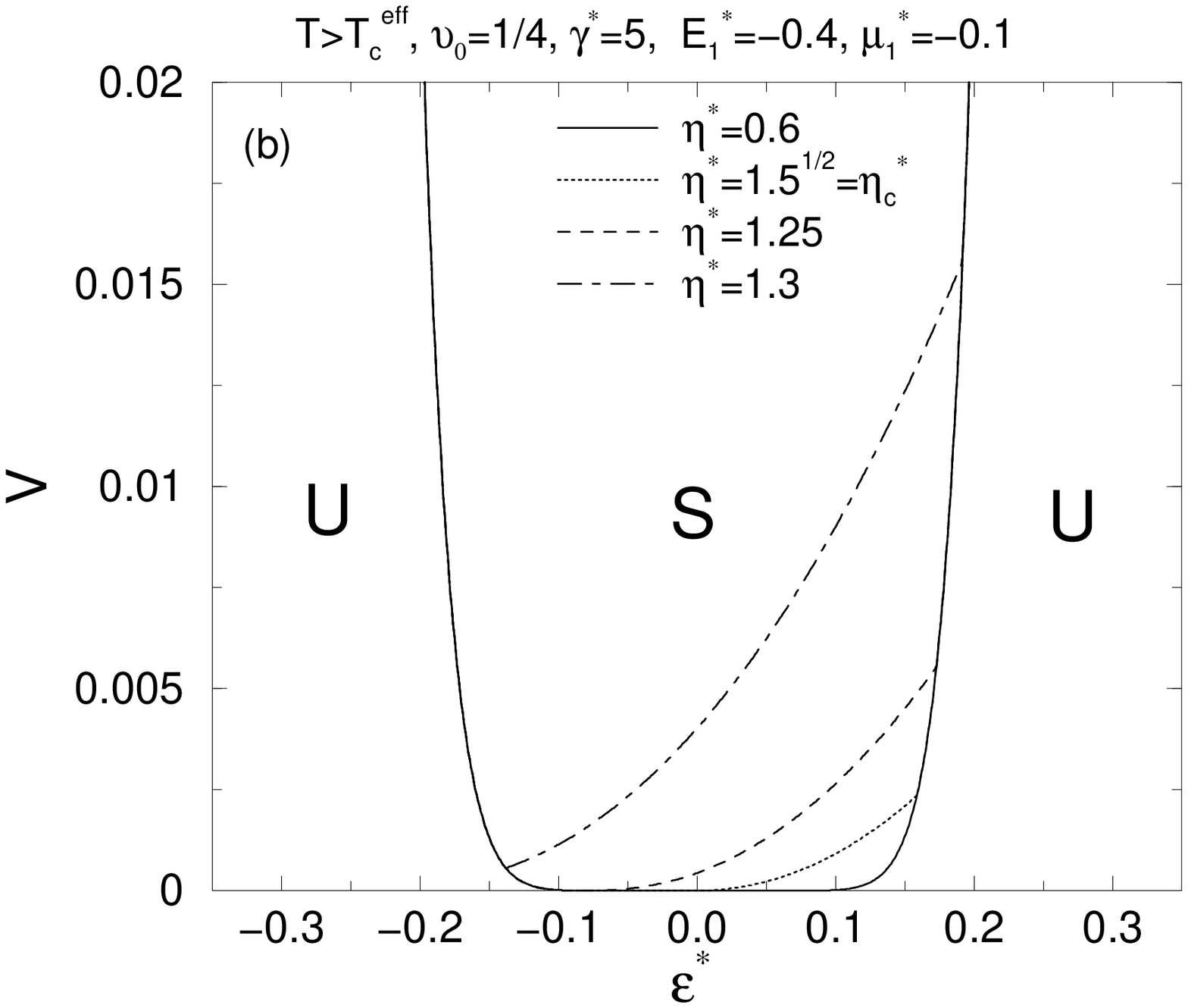}
\hskip 1.5cm
\includegraphics{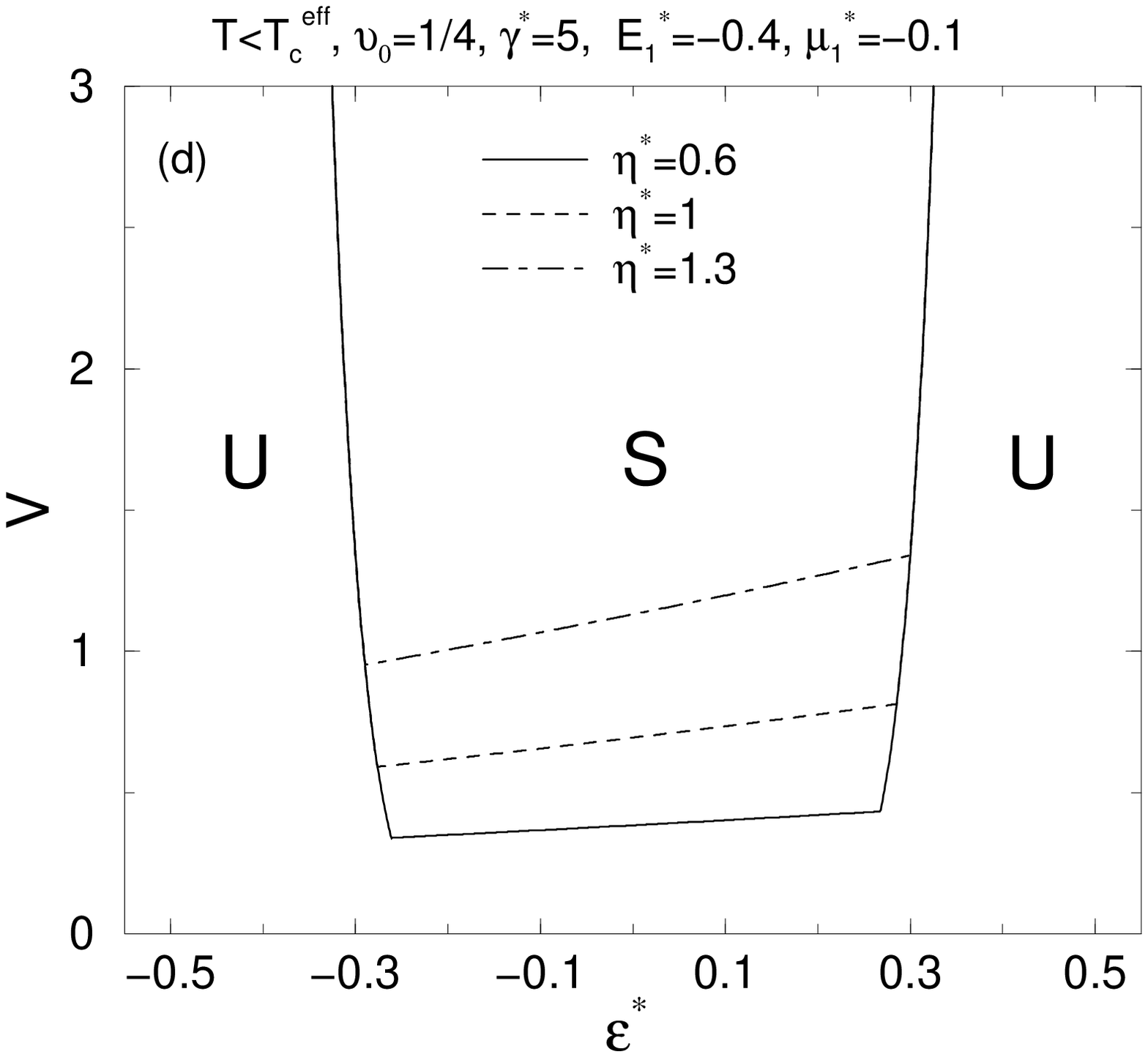}} 
\caption{\label{fig-epsV-SiGe}Stability diagrams of 
deposition rate $V$ versus misfit $\epsilon^*$, with SiGe-like
parameters $\nu_0=1/4$, $\gamma^*=5$, and different cases of
composition independent ((a) and (c): $E_1^*=\mu_1^*=0$) or 
composition dependent ((b) and (d): $E_1^*=-0.4$, $\mu_1^*=-0.1$) 
elastic moduli. (a) and (b) correspond to $T>T_c^{\rm eff}$ 
systems, while (c) and (d) are for $T<T_c^{\rm eff}$. Stable
and unstable regions are marked as ``S'' and ``U'', respectively.
$\eta^*$ values increase from bottom to top stability boundary
lines, as indicated in the figures, and in figure (a) all curves 
with $\eta^* \leq \eta_c^* =1.5^{1/2}$ cannot be distinguished 
from each other for the scale plotted here. Note that the vertical
scales of $V$ in (c) and (d) for $T<T_c^{\rm eff}$ are much larger 
than those of (a) and (b) for $T>T_c^{\rm eff}$.}
\end{figure}

\begin{figure}
\resizebox{0.55\textwidth}{!}{
\includegraphics{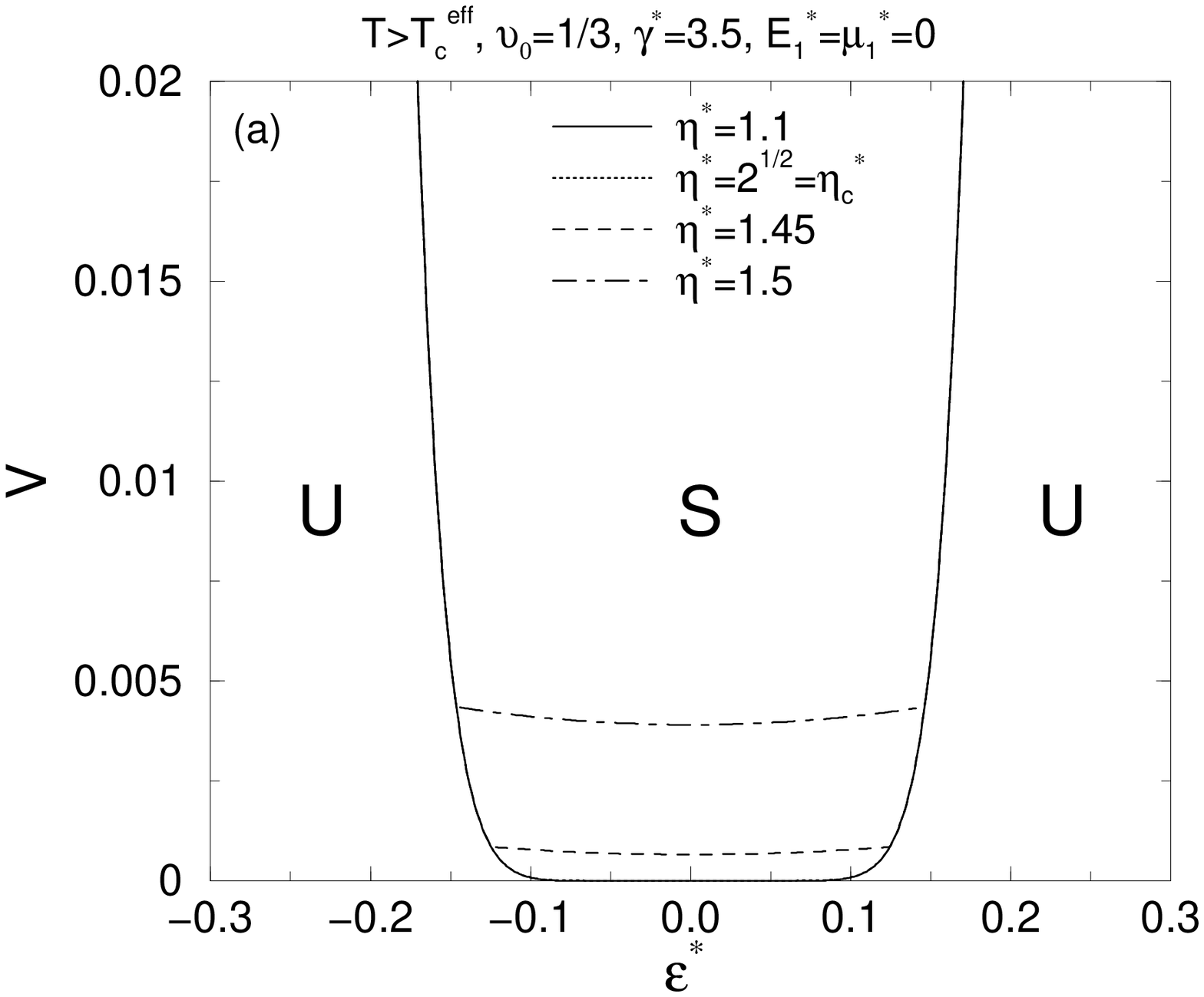}} \\
\resizebox{0.55\textwidth}{!}{
\includegraphics{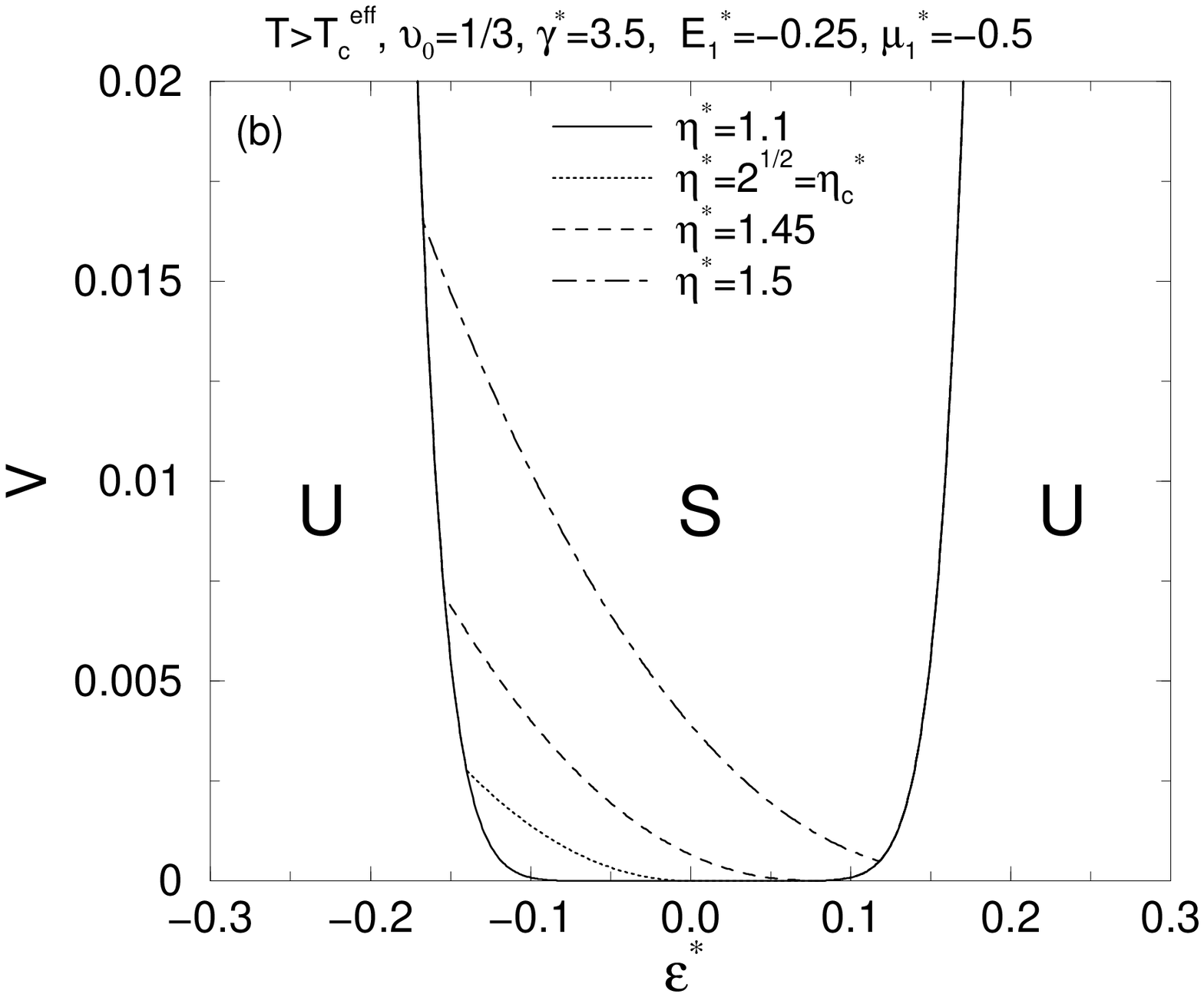}} 
\caption{\label{fig-epsV-InGaAs}Stability diagrams of deposition rate 
$V$ versus misfit $\epsilon^*$ similar to Fig. \ref{fig-epsV-SiGe}, 
except for $T>T_c^{\rm eff}$ and for InGaAs-like parameters $\nu_0=1/3$, 
$\gamma^*=3.5$, $\eta_c^*=2^{1/2}$, as well as $E_1^*=-0.25$ and 
$\mu_1^*=-0.5$. Compared with Fig. \ref{fig-epsV-SiGe}, different 
asymmetry with respect to sign of misfit $\epsilon^*$ is shown for (b), 
due to different material parameters. The diagrams for $T<T_c^{\rm eff}$
(not shown here) are similar to those of Fig. \ref{fig-epsV-SiGe}, except
the opposite asymmetry for nonzero $E_1^*$ and $\mu_1^*$.}
\end{figure}

In Figs. \ref{fig-epsV-SiGe} and \ref{fig-epsV-InGaAs} we plot 
the stability diagrams of dimensionless deposition rate $V$ 
versus misfit strain $\epsilon^*$ for different values of 
$\eta^*$, including different cases of growth temperatures 
(above or below coherent spinodal $T_c^{\rm eff}$) and of 
composition dependence or independence of elastic moduli. 
When one compares Figs. \ref{fig-epsV-SiGe} (a) and 
\ref{fig-epsV-InGaAs} (a), the stability diagrams are
qualitatively similar. For $E_1^* = \mu_1^* =0$, the 
stability boundaries are symmetric with respect to sign of 
$\epsilon^*$ and qualitatively insensitive to changes in 
$\nu_0$ or $\gamma^*$. The major qualitative difference 
between these two figures is seen for the case of composition 
dependent elastic constants, \textit{i.e.} when $E_1^*\neq 0$ 
and $\mu_1^*\neq 0$, as we compare Fig. \ref{fig-epsV-SiGe} 
(b) with Fig. \ref{fig-epsV-InGaAs} (b) which are for 
$T>T_c^{\rm eff}$. These stability results are clearly 
asymmetric with respect to the sign of misfit $\epsilon^*$. 
Similar asymmetry also occurs for $T<T_c^{\rm eff}$. This 
asymmetry emerges from the characteristic equation 
(\ref{eq-sigma}), and is enhanced by larger solute stresses 
(\textit{i.e.} larger values of $\eta^*$). However, the form 
of asymmetry is dependent on the choice of parameters. In Fig. 
\ref{fig-epsV-SiGe}, the parameters of SiGe-type are used and 
the stability diagrams show that the instability region of a 
film under compressive strain ($\epsilon^* >0$) is larger than 
that under tensile strain ($\epsilon^* <0$), while in Fig. 
\ref{fig-epsV-InGaAs} with InGaAs-type parameters, we have 
opposite result that the compressive films have larger stability 
region. This asymmetry of stability has also been found in earlier 
works of Guyer and Voorhees \cite{guyer1,guyer2,guyer4,guyer5} 
and Leonard and Desai \cite{leonard1,leonard3} (for the case of 
composition dependent Young modulus) when single surface mobility 
is considered. However, in these results only one kind of 
asymmetry is obtained, that is, tensile films are more stable than 
compressive films, while our calculations yields different kinds
of asymmetry due to different parameters $E_1^*$ and $\mu_1^*$
determined by composition dependent behavior of elastic moduli.
Note that most experiments are carried out at temperature
$T>T_c^{\rm eff}$, with growth parameters corresponding to
very small rescaled deposition rate $V$ and for rescaled $\eta^*$ 
($\sim 0.6$ for SiGe film and $\sim 1.1$ for InGaAs growth)
that is not large. Thus, to compare with available experimental 
observations, we should study the bottom part and the lowest
curve of diagrams in Figs. \ref{fig-epsV-SiGe} (b) and
\ref{fig-epsV-InGaAs} (b), which, however, have very weak 
asymmetry. In the next section, we will present 
the results for kinetic critical thickness on the scales 
corresponding to experimental parameters, and the asymmetry of 
(effective) stability will be shown more clearly, with more 
complex behavior for InGaAs-type films.

A common feature of the diagrams in Figs. \ref{fig-epsV-SiGe}
and \ref{fig-epsV-InGaAs} is that the system can be completely
stabilized by large enough deposition rate $V$, the value of 
which is determined by material parameters like $\epsilon^*$ 
and $\eta^*$. This stabilization effect at large deposition 
rate can be understood as follows. First, when the materials 
are deposited fast enough, the former surface layer is buried
and then frozen (due to miniscule bulk mobility) before 
the surface morphological undulation and phase segregation
have enough time to develop through surface diffusion.
Second, as described in dynamical equation (\ref{eq-phi}),
the deposition flux is uniform and then have the tendency 
to drive the surface composition profile to the homogeneous 
phase of incident flux. Note that in Fig. \ref{fig-epsV-SiGe}
the scale of $V$ in the stability diagrams ((c) and (d)) for 
$T<T_c^{\rm eff}$ is much larger than those ((a) and (b)) for 
$T>T_c^{\rm eff}$, due to the fact that in usual coherent systems, 
spinodal decomposition is apt to occur below the temperature 
$T_c^{\rm eff}$, and then faster deposition is needed to suppress 
it. This effect of stabilization at high deposition rates has 
been found by Guyer and Voorhees \cite{guyer2,guyer4,guyer5} and 
Leonard and Desai \cite{leonard1,leonard3}, but is absent in 
recent work of Spencer \textit{et al}. \cite{spencer1,spencer3} 
for the case of equal mobilities of alloy components. However, 
in the results of Guyer and Voorhees, the stress driven 
instabilities can be suppressed only for films under tensile 
strain ($\epsilon^* <0$), while for compressive layers this 
stabilization by high $V$ is not possible and then systems are 
always unstable. This is different from our results where films 
of both $\epsilon^*>0$ and $\epsilon^*<0$ can be stabilized for 
symmetric or asymmetric cases, as shown in Figs. \ref{fig-epsV-SiGe} 
and \ref{fig-epsV-InGaAs}. 

Compared to single-component film, one of the additional and 
crucial factors for alloy film stability is $\eta$, reflecting 
the compositional stress caused by different atomic sizes of 
alloy components. From the model of Guyer and Voorhees 
\cite{guyer1,guyer2,guyer4,guyer5} it is found that the
compositional stress can stabilize the film modulations.
However, recent study of Spencer \textit{et al.} \cite{spencer1,%
spencer2,spencer3} shows that when one considers single atomic 
mobility, the coupling of solute strain ($\eta$) and film-substrate 
misfit strain $\epsilon$ always enhances the instability and makes 
the alloy film more unstable than the single-component film, and 
only for the case of unequal surface mobilities can the solute 
strain lead to a decrease of the film instability with a possibility 
of stabilizing the system through deposition. Here we have used one 
effective surface mobility to describe the dynamics of the alloy 
film, but have more complicated and different results for the role 
of solute stress (or $\eta$). To obtain the effect of $\eta$ more 
clearly, without the interplay of composition dependence of 
elastic constants, we consider the case of $E_1^* = \mu_1 ^* =0$, 
corresponding to Figs. \ref{fig-epsV-SiGe} (a), (c) and 
\ref{fig-epsV-InGaAs} (a). These three diagrams show that
the joint morphological and compositional instabilities can
be suppressed by large deposition rate, as discussed above,
resulting in less unstable film compared with single-component
system where the Asaro-Tiller-Grinfeld morphological instability 
(as described in Eq. (\ref{eq-sig-h2})) always occurs for any 
deposition rate. In this sense our result is in agreement with
that of Guyer and Voorhees. However, in the presence of solute
stress the diagrams in Figs. \ref{fig-epsV-SiGe} and 
\ref{fig-epsV-InGaAs} exhibit the smaller stable regions for
larger $\eta^*$, corresponding to the destabilizing effect of 
large compositional strain, which is opposite to the effect
found by Guyer and Voorhees but similar to that of Spencer 
\textit{et al.}. A new feature of our results is in the 
stability diagrams for $T>T_c^{\rm eff}$, \textit{i.e.} Figs. 
\ref{fig-epsV-SiGe} (a) and \ref{fig-epsV-InGaAs} (a), where
the system is always stable for small misfits $|\epsilon^*|
< \epsilon_0^*$ when $\eta^* < \eta_c^*$, with $\eta_c^*$ 
defined in Eq. (\ref{eq-st-eps0+}) and $\epsilon_0^*$ dependent 
on the value of $\eta^*$. Thus, only when the compositional 
stress exceeds a critical value, \textit{i.e.} $\eta^*>\eta_c^*$, 
can the instability of the film be significantly enhanced. 
This is different from all the previous results, and can 
be explained by Eq. (\ref{eq-st-eps0+}) in the limit of zero 
misfit, which shows analytically that a film is stable against 
any disturbance when $\eta^*<\eta_c^*$. This feature does not
exist for $T<T_c^{\rm eff}$, which can also be expected from
the stability condition (\ref{eq-st-eps0-}) for $\epsilon^*=0$.

\section{\label{sec:VI}Kinetic Critical Thickness}

According to the stability calculations shown above, in general
the morphological and compositional modes are coupled, that is,
instability in one also implies instability in the other, and 
then the joint stability or instability is determined by diagrams 
illustrated in Figs. \ref{fig-epsV-SiGe} and \ref{fig-epsV-InGaAs}. 
However, in some experiments, although the chosen material parameters 
and experimental conditions correspond to the unstable region of our 
theoretical diagram, the film profiles are in fact observed to be 
stable. This may be due to the kinetic stabilization effect of growing 
film, as proposed by Spencer \textit{et al.}.\cite{spencer4,spencer5}
When the real part of perturbation growth rate $\sigma$ is positive,
the instability develops and grows with time. However, at early
time of growth the surface perturbations have not enough time to
develop and become observable compared to constant deposition of
materials flux. Only after a sufficient time, that is, when the 
film is thick enough, can the perturbations develop substantially
and be apparent relative to the planar basic state of the film.
Then the perturbation growth rate ($\sigma$) can exceed the 
relative growth rate of the film basic state ($v/\bar{h}$). Thus, 
this competition between deposition and perturbation growth results 
in a kinetic critical thickness $h_c$, with the nondimensional
form given by 
\begin{equation}
h_c^* = V/\sigma
\label{eq-hc}
\end{equation}
in linear analysis, \cite{spencer1,spencer4,spencer5} where $\sigma$ 
is usually approximated by the maximum perturbation growth rate 
$\sigma_{\rm max}$ with respect to different growing mode $k$. 
Below this critical thickness $h_c$, the instability of the growing 
film is effectively suppressed by deposition flux and may not be
seen experimentally, while above $h_c$ the surface undulations are
seen. This phenomenon for the onset of stress driven instability
at certain film thickness has been observed by experiments, for
both the systems of SiGe \cite{perovic,tromp} and InGaAs. \cite{%
gendry,grandjean,chokshi}

\begin{figure}
\resizebox{0.59\textwidth}{!}{
\includegraphics{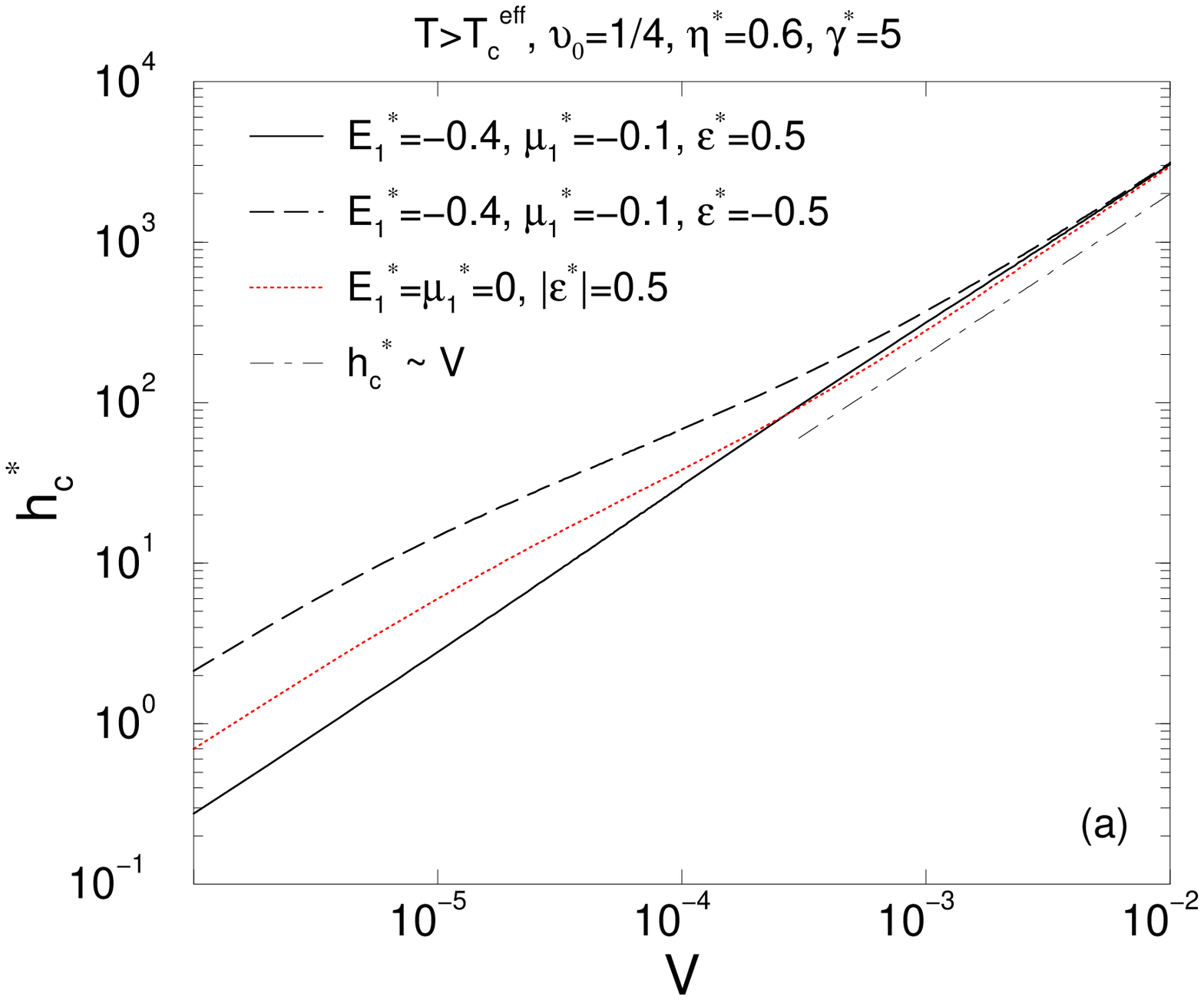}} \\
\resizebox{0.59\textwidth}{!}{
\includegraphics{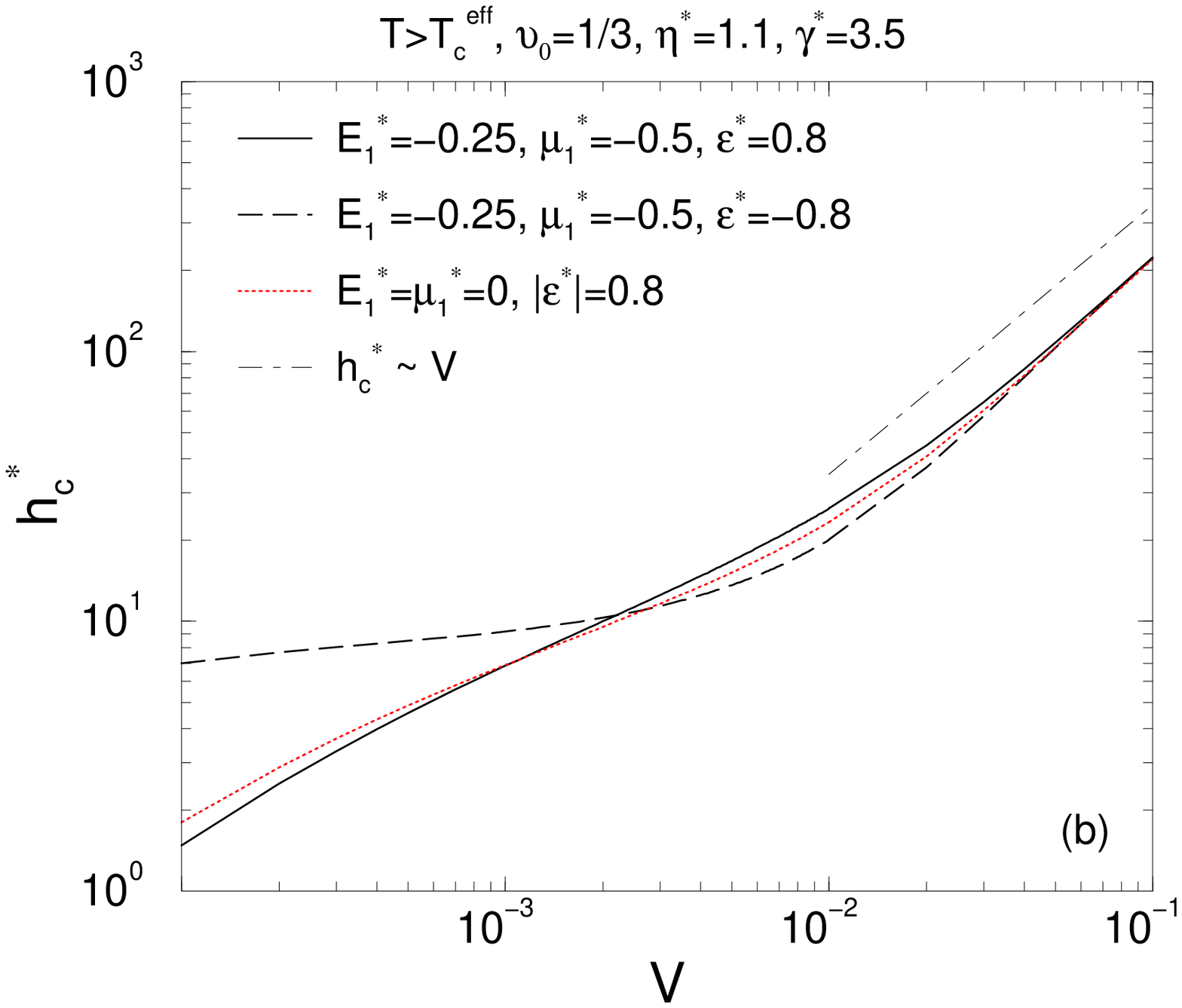}} 
\caption{\label{fig-hc-V}Kinetic critical thickness (nondimensional)
$h_c^*$ as a function of rescaled deposition rate $V$ for growth
temperature above $T_c^{\rm eff}$. The solid ($\epsilon^*>0$) and
dashed ($\epsilon^*<0$) curves are for the case of composition
dependent elastic moduli, while the dotted curve represents the 
result for composition independent moduli. The dot-dashed line of 
linear relationship $h_c^* \propto V$ is also drawn for comparison.
(a) The material parameters are chosen to qualitatively represent 
the SiGe-like film: $\nu_0=1/4$, $\eta^*=0.6$, $\gamma^*=5$, 
$E_1^*=-0.4$, $\mu_1^*=-0.1$, and $|\epsilon^*|=0.5$; (b) The 
parameters are expected to qualitatively represent the InGaAs-like 
alloy: $\nu_0=1/3$, $\eta^*=1.1$, $\gamma^*=3.5$, $E_1^*=-0.25$, 
$\mu_1^*=-0.5$, and $|\epsilon^*|=0.8$.}
\end{figure}

>From the mechanism of kinetic stabilization described above, one
can expect that kinetic critical thickness $h_c^*$ will increase
with deposition rate, which is verified by our calculations of
$h_c^*$ versus $V$ shown in Fig. \ref{fig-hc-V} for growth 
temperature $T>T_c^{\rm eff}$. For both the SiGe-type and 
InGaAs-type parameters, the log-log plots in Figs. \ref{fig-hc-V} 
(a) and (b) show that when the rescaled deposition rate $V$ is 
not large (corresponding to most experimental cases), generally 
the relationship between $h_c^*$ and $V$ is not linear, and the 
nature of the rise of $h_c^*$ with $V$ for compressive and tensile 
misfits is different for the case of composition dependent elastic 
moduli. At sufficiently high deposition rate, the curves of $h_c^*$
for opposite signs of misfit $\epsilon^*$ and for different 
composition dependent forms of elastic moduli converge, and its 
variation with respect to deposition rate is linear, which can be 
seen on scale of $V$ larger than that of Fig. \ref{fig-hc-V} but 
is not shown here. These properties imply that the maximum of 
perturbation growth rate $\sigma$ depends on small deposition 
rate $V$ as well as on the values of $E_1^*$ and $\mu_1^*$, but 
is independent of them when $V$ is large enough. 

>From the plots in Fig. \ref{fig-hc-V} we can also obtain the 
results of stability asymmetry with respect to the sign of misfit,
except for large enough deposition rate $V$ which results in
the presence of symmetry of $h_c$ in $\epsilon$. For SiGe-type 
system with composition dependent elastic moduli, Fig. \ref{fig-hc-V} 
(a) shows that kinetic critical thickness for compressive films 
($\epsilon^* >0$) is smaller than that of tensile films ($\epsilon^* 
<0$) when $V$ is not large enough, corresponding to the phenomenon 
that the growing films under tension are effectively more stable, 
as observed in SiGe experiments. \cite{xie1} The asymmetric property 
for InGaAs-type system is more complicated, that is, it depends on 
the deposition rate for $T>T_c^{\rm eff}$, as shown in Fig. 
\ref{fig-hc-V} (b). For intermediate values of $V$, the compressive
films have larger $h_c^*$ compared to the films under tension, 
making compressive films more stable. For small $V$, $h_c^*$ of 
compressive films is smaller, leading to less effective stability
which is similar to SiGe-type films. Note that the asymmetric 
property obtained here for InGaAs is different from the result 
of higher $\eta^*$ values shown in the stability diagram of Fig. 
\ref{fig-epsV-InGaAs} (b), which in fact represents the asymmetry 
of absolute stability, but not that of effectively kinetic 
stabilization discussed here.

This deposition dependent behavior of asymmetry for InGaAs films
can qualitatively explain the contradiction in recent experimental
observations. In the experiment of Okada \textit{et al.},\cite{okada} 
compressively strained layers are found to be more stable than tensile 
layers, whereas the observation of Guyer \textit{et al.} \cite{guyer3} 
yields opposite result. According to our calculations here, this 
inconsistency is attributed to different deposition rate used in 
these experiments: The growth conducted by Guyer \textit{et al.} has 
much lower deposition rate and then corresponds to the asymmetry 
similar to SiGe films but opposite to that of Okata \textit{et al.}. 
The direct measurement of critical thickness for the onset of 
instability has been carried out by Gendry \textit{et al.} 
\cite{gendry} for different signs of misfit and different deposition 
rate. Since our calculations mainly focus on 50-50 mixture, different 
from the alloy concentrations used in Ref. \onlinecite{gendry}, here 
we only compare the theoretical and experimental results qualitatively. 
As observed during MBE growth of Gendry \textit{et al.}, for the 
undoped layers with As-stabilized surface showing strong ($2\times 4$) 
reconstruction, the critical thickness in compression is found to
be less than that in tension when deposition rate is small, while
for high deposition rate, larger $h_c$ in compression is obtained.
This is consistent with our findings in Fig. \ref{fig-hc-V} (b)
for the case of composition dependent elastic moduli.

\begin{figure}
\resizebox{0.6\textwidth}{!}{
\includegraphics{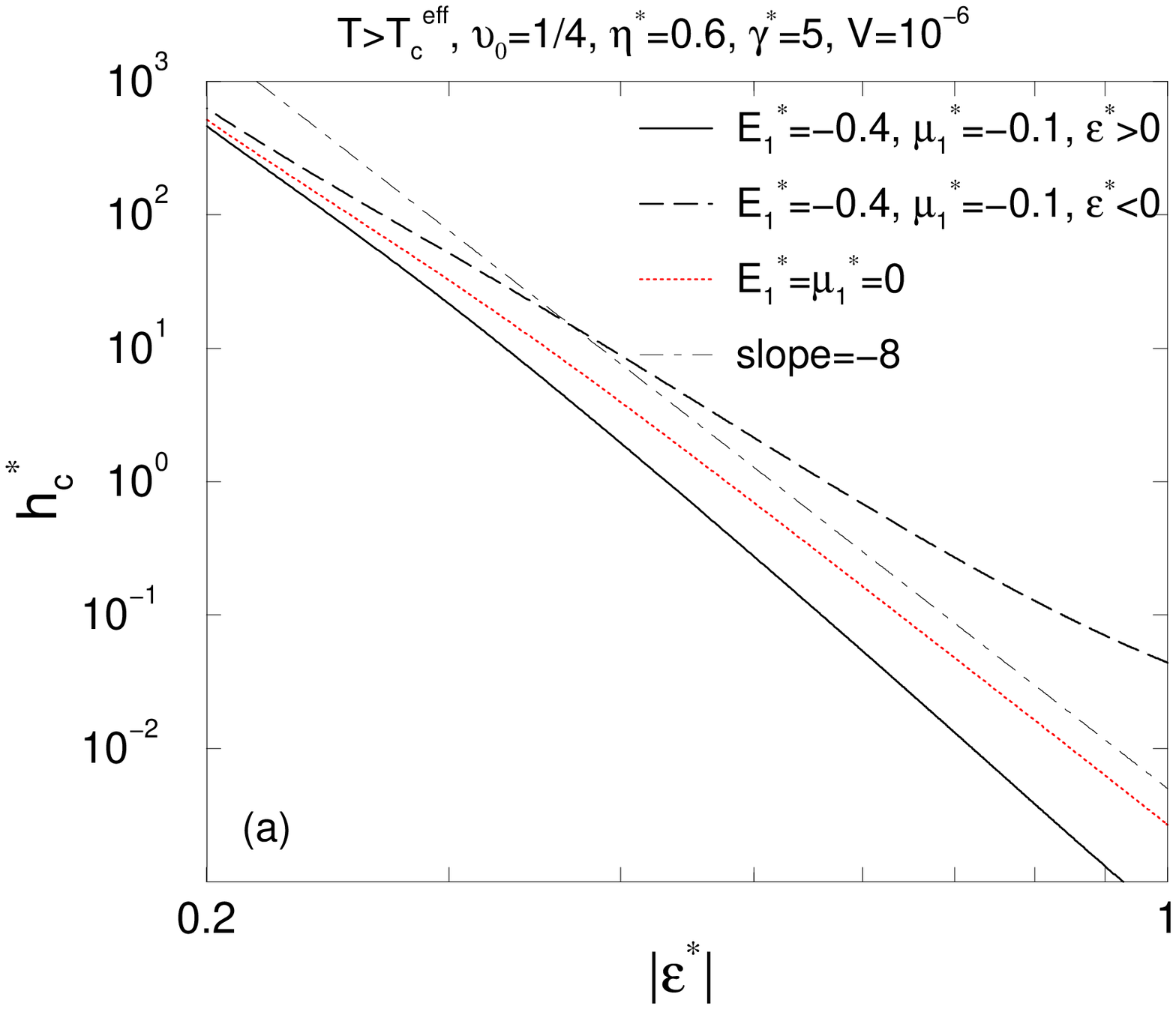}} \\
\resizebox{0.6\textwidth}{!}{
\includegraphics{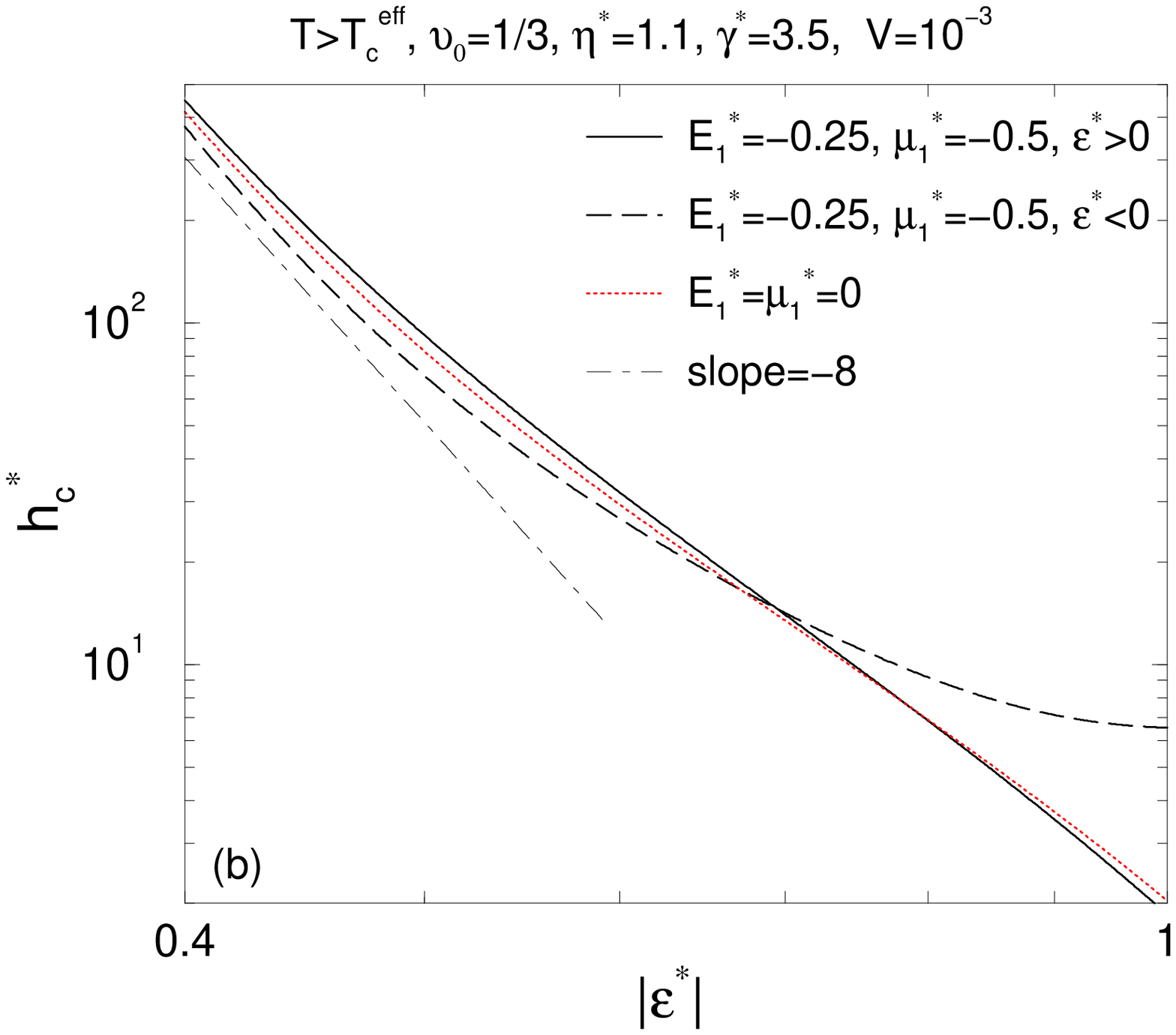}} 
\caption{\label{fig-hc-eps}Rescaled kinetic critical thickness 
$h_c^*$ as a function of absolute value of rescaled misfit strain 
$|\epsilon^*|$ for temperature $T>T_c^{\rm eff}$. Similar to Fig. 
\ref{fig-hc-V}, the solid and dashed curves are for positive and 
negative misfit $\epsilon^*$ respectively, with composition
dependent elastic moduli, while the dotted curve corresponds to
the case of composition independent moduli. The dot-dashed line
represents the power law scaling behavior with exponent $-8$,
which is used for comparison. In (a), the rescaled deposition
rate is chosen as $V=10^{-6}$, and in (b) it is $V=10^{-3}$.
All the other paramters are the same as those in Fig. \ref{fig-hc-V}.
}
\end{figure}

The scaling behavior for critical thickness $h_c$ with respect to
misfit $\epsilon$ has also attracted attention in recent work.
Since the system becomes more unstable with the increase of misfit
strain, $h_c$ is expected to decrease with larger $\epsilon$.
For single-component, lattice mismatched films, where the effect 
of solute stress is absent, the kinetic critical thickness is
found to follow the $\epsilon^{-8}$ power law. \cite{spencer4,%
spencer5} Our calcuations also lead to $\epsilon^{-8}$ power law 
as follows: From Eq. (\ref{eq-hc}), $h_c^* = V/\sigma_{\rm max}$;
from Eq. (\ref{eq-sig-h2}), one can show that the maximum growth
rate occurs at $k_{\rm max} = 3{\epsilon^*}^2/4\gamma^*$ which
leads to $h_c^* = (256{\gamma^*}^3 V)/(27{\epsilon^*}^8)$.
The observed behavior for multi-component system like SiGe is 
different. The experiment carried out by Perovi\'{c} 
\textit{et al.} \cite{perovic} finds that kinetic critical 
thickness $h_c$ for coherent SiGe/Si(100) varies with misfit as 
$\epsilon^{-4}$, while recent observation of Tromp \textit{et al.}
\cite{tromp} found a much slower dependence $\epsilon^{-1}$.
Our results for the dependence of rescaled critical thickness
$h_c^*$ on $\epsilon^*$ are shown in Fig. \ref{fig-hc-eps},
and indicate that generally the scaling of $h_c^*$ versus
$\epsilon^*$ is not a power law. For the SiGe-type system,
as illustrated in Fig. \ref{fig-hc-eps} (a), the curve (dotted)
for the case of composition independent elastic moduli ($E_1^*=
\mu_1^*=0$) presents a behavior close to $1/{\epsilon^*}^8$ for
large misfit $\epsilon^*$, but has slower decreasing form for
small $\epsilon^*$. When we consider the composition dependence 
of elastic constants ($E_1^* \neq 0$ and $\mu_1^* \neq 0$), the 
decrease of $h_c^*$ with $\epsilon^*$ is slightly faster than 
${\epsilon^*}^{-8}$ for the compressive layers ($\epsilon^*>0$),
but much slower for the tensile films. In Fig. \ref{fig-hc-eps} 
(b), corresponding to the InGaAs-type case, $h_c^*$ scales 
similar to ${\epsilon^*}^{-8}$ only for small misfit $\epsilon^*$, 
but decreases much slower at larger values of misfit, in particular 
for films grown under tension if $E_1^* \neq 0$ and $\mu_1^*\neq 0$.

The asymmetry of effective stability (\textit{i.e.} thickness
$h_c$) with respect to the sign of misfit can also be seen in
Fig. \ref{fig-hc-eps} for the case of composition dependent 
elastic moduli. For the SiGe-type system (Fig. \ref{fig-hc-eps} 
(a)) $h_c^*$ for tensile films is always larger than that of 
compressive films, that is, tensile layers are effectively more 
stable, as found earlier. However, the InGaAs-type system yields 
more complicated scaling behavior. For large misfit $\epsilon^*$,
Fig. \ref{fig-hc-eps} (b) exhibits that films under tension have 
higher $h_c^*$, and then can be observed as more stable, similar 
to the result of Fig. \ref{fig-hc-eps} (a). This is in accord with 
the result of $h_c^*$ versus $V$ shown in Fig. \ref{fig-hc-V} (b), 
since here the rescaled deposition rate $V$ is chosen as $10^{-3}$, 
corresponding to the small $V$ region of Fig. \ref{fig-hc-V} (b)
where one has large $|\epsilon^*|=0.8$ and SiGe-like behavior of
asymmetry. For intermediate values of $|\epsilon^*|$, asymmetry
opposite to SiGe-type system is obtained, that is, the $h_c^*$
curve for compressive layers lies above that for tensile layers,
resulting in more stable compressive films. When $|\epsilon^*|$ 
gets smaller (not shown here), the difference of $h_c^*$ values 
between $\epsilon^* > 0$ and $\epsilon^* < 0$ becomes negligible, 
and then we have the symmetry of stability for films under 
compression and under tension even when elastic constants depend 
on composition field. Therefore, combining Figs. \ref{fig-hc-V} 
(b) and \ref{fig-hc-eps} (b), we conclude that the asymmetry of 
effective stabilization for InGaAs-type films is dependent on 
both the deposition rate and the magnitude of misfit strain.

\section{\label{sec:VII}Conclusions}

In this paper, we have determined the stability properties for 
growth of dislocation-free and coherent, strained alloy films by 
developing a continuum dynamical model and performing a linear 
stability analysis. In our model there are two important and new
features, the coupling between top surface of the film and the 
bulk underneath, and the composition dependence of elastic moduli 
($E$ and $\mu$), which result in new stability results. Both the 
thermodynamics and elastic effects are considered. Besides 
the phase separation mechanism at low temperature and the role 
of surface energy that favours the planar surface, we also include
the effects of elasticity which can arise due to (i) the mismatch 
of lattice constant between a growing film and the substrate, 
resulting in nonzero misfit strain $\epsilon$; (ii) the dependence 
of lattice constant on composition, leading to nonzero solute 
strain represented by a coefficient $\eta$; and (iii) the dependence 
of elastic moduli on composition, as given in Eq. (\ref{eq-Eu}). 
Moreover, the deposition rate can play an important role for 
the growing films. The main results, due to the interplay of 
these factors, that we obtain can be summarized as follows.

For single-component epitaxial growth, when one studies the 
stability of the free surface of a growing film, the morphological 
corrugation may lead to the formation of coherent islands. This 
occurs on account of the misfit strain caused by lattice mismatch.  
In contrast, for multi-component growth like MBE, the time evolution 
of the morphological profile is coupled to the dynamics of composition. 
Physically the coupling occurs since smaller (larger) atoms of 
the film prefer to incorporate in regions of compressive (tensile) 
stress. This leads smaller atoms to preferentially diffuse to troughs 
(peaks) and larger atoms to segregate to peaks (troughs), for a film 
under compression (tension). As a result of the coupling, there 
is a joint modulation of the surface morphological profile and the 
alloy composition, as well as a common perturbation growth rate 
$\sigma$ (see Eqs. (\ref{eq-h1}), (\ref{eq-phi1}), and 
(\ref{eq-sigma})). Thus a planar surface can be stabilized 
only if alloy decomposition is suppressed and vice versa. Some 
experiments (SiGe \cite{walther} and InGaAs \cite{okada,peiro,%
gonzalez}) have verified such a simultaneous development of 
morphological and compositional modulations.

For \textit{bulk} strained alloys, the compositional profile is 
homogeneous above the bulk coherent spinodal temperature 
$T_c^{\rm eff}$, and the decomposition can occur only below 
$T_c^{\rm eff}$. However, for multi-component epitaxial \textit{films}, 
alloy decomposition occurs both theoretically and experimentally 
even for temperature $T$ higher than $T_c^{\rm eff}$. This is due 
to two major effects. First, due to the coupling between
morphological and compositional profiles, the phase separation can 
be driven by the difference in strain energy densities along the 
surface caused by the morphological undulation. Even in the absence 
of $\eta$, that is, when the atomic size difference is negligible, 
the composition dependence of the film elastic constants can be
the cause of coupling between morphology $h$ and composition $\phi$, 
as shown in Eqs. (\ref{eq-quad}) and (\ref{eq-a10-1}) for special 
case of $\epsilon^* \neq 0$ but $\eta^*=0$. The second effect is 
related to solute stress (or $\eta$) itself. As depicted in Eqs.
(\ref{eq-a10-2}) and (\ref{eq-st-eps0+}) for $\epsilon^* = 0$
and $\eta^* \neq 0$, when the morphological and compositional
degrees of freedom are decoupled and we have stable surface
morphology, alloy decomposition can also occur for $T>T_c^{\rm eff}$ 
if $\eta^*$ is large enough. This means that when the solute 
stress is too strong, corresponding to very large difference in 
the atomic sizes of alloy components, the small and large species
prefer to segregate with respect to very small background 
disturbances of surface profile.

This compositional instability, as well as the morphological 
modulation, can be completely suppressed by high deposition rate,
both for $T$ above and below coherent spinodal $T_c^{\rm eff}$, 
as seen in the stability diagrams of Figs. \ref{fig-epsV-SiGe} 
and \ref{fig-epsV-InGaAs} as well as from the analytic results 
derived for special cases in Sec. \ref{sec:V-A}. Generally, 
increasing the magnitude of misfit $\epsilon$ increases the 
instability, and larger solute strain ($\eta^*$) also enhances 
the instability for most of the cases. Furthermore, the nonphysical 
short wavelength instability is absent in our work, due to the 
consideration of gradient free energy.

Including the coupling between film surface and underlying bulk
has resulted in some interesting results. One of the new features is
on the role of $\eta$. Our analysis naturally brings a critical 
value $\eta_c^*$ for the stability at temperature $T>T_c^{\rm eff}$.
For $|\eta^*|$ below $\eta_c^*$, the system is always stable within 
a range of misfit $\epsilon^*$, regardless of any deposition rate 
$V$. Only for $|\eta^*| > \eta_c^*$ can the increase of $\eta^*$
significantly enhance the instability; this is shown in Eq. 
(\ref{eq-st-eps0+}) and Fig. \ref{fig-eps0} for $\epsilon^* = 0$,
as well as in Figs. \ref{fig-epsV-SiGe} and \ref{fig-epsV-InGaAs}
for nonzero $\epsilon^*$. When the surface-bulk coupling is 
neglected, the stability results do not have any critical $\eta^*$. 
Our results for the scaling behavior of kinetic critical thickness 
$h_c$ (the growing film is effectively stabilized by deposition 
up to thickness $h_c$) are also different from earlier work. In 
general, the dependence of nondimensional critical thickness 
$h_c^*$ on rescaled deposition rate $V$ is not linear for 
intermediate and small $V$ (corresponding to most experimental 
cases), and its dependence on rescaled misfit $\epsilon^*$ is 
not necessarily a power law, as shown in Figs. \ref{fig-hc-V} 
and \ref{fig-hc-eps} respectively, even for $E_1^* = \mu_1^* =0$.

The asymmetry of stability with respect to the sign of misfit 
$\epsilon^*$ is found in the systems with composition dependent 
film elastic moduli, while for the composition 
independent case ($E_1^*=\mu_1^*=0$), it is absent. 
Figures \ref{fig-epsV-SiGe} and \ref{fig-epsV-InGaAs}
show the diagrams for absolute stability. For SiGe-type system,
the films grown under tension ($\epsilon^*<0$) are expected to 
be more stable than those under compression ($\epsilon^*>0$), 
while the form of asymmetry is opposite for InGaAs-type films. 
For the effective stabilization, as measured by critical thickness 
$h_c$, the asymmetry can also be found unless the deposition rate 
is too large. According to our calculations, a tensile
film, with SiGe-type parameters, will be observed as stable up to 
a thickness larger than that for a compressive film, as shown 
in Figs. \ref{fig-hc-V} (a) and \ref{fig-hc-eps} (a). However, for 
InGaAs-type system the asymmetry depends on both the deposition
rate and the magnitude of misfit. As found in Fig. \ref{fig-hc-V}
(b), the system will have asymmetry similar to SiGe-type at small
$V$, but for intermediate values of $V$ reverse asymmetry is 
obtained, and then the compressive films are effectively more stable. 
This can explain the observations of recent experiments on InGaAs 
epitaxial films. \cite{okada,guyer3,gendry} When the deposition rate 
is fixed, the forms of asymmetry is opposite for large and intermediate
magnitude of misfit $\epsilon^*$, but at small misfit the symmetry
with respect to the sign of $\epsilon^*$ is obtained (see Fig.
\ref{fig-hc-eps} (b)).

As discussed above and shown in Figs. \ref{fig-eps0}--\ref{fig-hc-eps},
the stability properties and asymmetry depend on the chosen material 
system and the experimental conditions (such as deposition rate $v$ 
and growth temperature $T$). E.g., if we desire a system with very 
strong modulations, stability diagrams (Figs. \ref{fig-epsV-SiGe} and 
\ref{fig-epsV-InGaAs}) and the variation of kinetic critical thickness 
(Figs. \ref{fig-hc-V} and \ref{fig-hc-eps}) show that the instability
requires large misfit $\epsilon$, large solute coefficient $\eta$, 
small deposition rate $v$, and large enough film thickness. Also 
note that the stability results presented here are with respect to 
dimensionless parameters $V$, $\epsilon^*$, $\eta^*$, and $\gamma^*$, 
which are all temperature dependent according to Eqs. (\ref{eq-scale}), 
(\ref{eq-l0}), and (\ref{eq-tau0}), via the coefficients $r$ and
$\Gamma_{\phi}$. The explicit temperature dependence of 
$\Gamma_{\phi}=\Gamma_{h}/\delta$ is given after Eq. (\ref{eq-h}).
Thus, even when we already choose the growing 
system with fixed material parameters, the stability behavior will
vary with growth temperature, through the variation of the dimensionless
parameters.

All of the above results show that the problem of multi-component 
epitaxial growth, which is essentially a problem of nonequilibrium, 
is rich in interesting physics. It involves competition between 
thermodynamics and deposition, and coupling between morphology 
and composition. More interesting and richer results can be 
obtained if one further studies the detailed patterns and structures 
of the growing film, beyond the early evolution regime considered 
in this paper.

\begin{acknowledgments}
This work was supported by the NSERC of Canada.
\end{acknowledgments}

\appendix

\section{\label{appendA}General results for mechanical 
equilibrium equation}

As discussed in Sec. \ref{sec:III}, the mechanical equilibrium
equation for the film is considered to zeroth order of elastic
constants, and then it is linear with the expansion form:
\begin{equation}
(1-2\nu^f_0)(\partial_z^2 -q^2) \left [
\begin{array}{c}
\hat{u}^f_x \\ \hat{u}^f_y \\ \hat{u}^f_z
\end{array} \right ] + \left [
\begin{array}{c}
iq_x \\ iq_y \\ \partial_z
\end{array} \right ] 
\left [ iq_x \hat{u}^f_x + iq_y \hat{u}^f_y + 
\partial_z \hat{u}^f_z - 2(1+\nu^f_0) \eta \hat{\phi} \right ]=0.
\label{eq-equi-q}
\end{equation}
The corresponding general solution is 
\begin{eqnarray}
\hat{u}^f_i &=& \left [
\begin{array}{c}
\alpha_x \\ \alpha_y \\ \alpha_z 
\end{array} \right ] \cosh(qz) + \left [
\begin{array}{c}
\beta_x \\ \beta_y \\ \beta_z
\end{array} \right ] \sinh(qz) \nonumber\\
&-& \left [
\begin{array}{c}
Ciq_x/q \\ Ciq_y/q \\ D
\end{array} \right ] z \sinh(qz) - \left [
\begin{array}{c}
Diq_x/q \\ Diq_y/q \\ C
\end{array} \right ] z \cosh(qz)
+ \left ( \frac{1+\nu^f_0}{1-\nu^f_0} \right ) \eta \left [
\begin{array}{c}
iq_x \hat{W} \\ iq_y \hat{W} \\ \partial_z \hat{W}
\end{array} \right ], 
\label{eq-u^f-q}
\end{eqnarray}
with $\hat{W}$ defined by Eq. (\ref{eq-W}) of Sec. \ref{sec:III}
and 
\begin{eqnarray}
C &=& \frac{1}{3-4\nu^f_0} (iq_x \alpha_x + iq_y \alpha_y
+q \beta_z), \nonumber\\
D &=& \frac{1}{3-4\nu^f_0} (iq_x \beta_x + iq_y \beta_y
+q \alpha_z).
\label{eq-CD}
\end{eqnarray}
The equation for the substrate displacement is similar to
Eq. (\ref{eq-equi-q}) but with $\hat{\phi}$ equal to zero.
Considering the boundary condition (\ref{eq-bound3-q}) for 
the decay of strains far from the interface, we have the
solution form
\begin{equation}
\hat{u}^s_i = \left [
\begin{array}{c}
u_x^0 \\ u_y^0 \\ u_z^0
\end{array} \right ] e^{qz} - \left [
\begin{array}{c}
iq_x/q \\ iq_y/q \\ 1
\end{array} \right ] Bze^{qz},
\label{eq-u^s-q}
\end{equation}
where
\begin{equation}
B=\frac{1}{3-4\nu^s} (iq_x u_x^0 + iq_y u_y^0 + q u_z^0).
\label{eq-B}
\end{equation}
The expression of stresses and strains can then be obtained
through the above solutions, e.g.,
\begin{eqnarray}
\hat{u}^f_{ll} &=& 2(1-2\nu^f_0) \left [ C \cosh(qz) 
+ D \sinh(qz) \right ] + \left ( \frac{1+\nu^f_0}{1-\nu^f_0} 
\right ) \eta \hat{\phi}, \nonumber\\
\hat{u}^f_{zz} &=& q \left [ \alpha_z \sinh(qz) + 
\beta_z \cosh(qz) \right ] - (D+Cqz) \sinh(qz) \nonumber\\
&-& (C+Dqz) \cosh(qz)  + \left ( \frac{1+\nu^f_0}{1-\nu^f_0} 
\right ) \eta \partial_z^2 \hat{W},
\label{eq-u_lz}
\end{eqnarray}
which are used in the elastic free energy calculation of
Sec. \ref{sec:IV}.

>From linearized boundary conditions (\ref{eq-bound1-q})--%
(\ref{eq-bound22-q}) at film surface and film-substrate interface,
as well as the consideration that both $\hat{\phi}$ and $\hat{W}$
are zero at $z=\bar{\zeta}=0$, we can determine the parameters 
$\alpha_i$, $\beta_i$, $u_i^0$, and then $C$, $D$, and $B$ in 
term of $\hat{h}$, $\hat{\zeta}$ and values $\hat{\phi}$ or 
$\hat{W}$ evaluated at $z=\bar{h}$. Here we omit the tedious 
deduction process and only present the results, which are:
\begin{eqnarray}
&iq_x\alpha_x + iq_y\alpha_y = & A^{-1} \rho e^f \left \{
q\bar{u}\hat{\zeta} \left [ -e^s e^f - (b^s - \rho c^s) 
(q\bar{h} - \sinh(q\bar{h})\cosh(q\bar{h}))
-e^s \sinh^2(q\bar{h}) \right ] \right. \nonumber\\
&& +\left [ q\bar{u}\hat{h} + \left ( \frac{1+\nu^f_0}
{1-\nu^f_0} \right ) \eta q \left ( \partial_z\hat{W}
\right )_{z=\bar{h}} \right ] 
\left [ e^s e^f\cosh(q\bar{h}) \right. \nonumber\\
&& \left. + 
(b^s b^f + \rho c^s)\sinh(q\bar{h}) + (b^s -\rho c^s)
q\bar{h}\cosh(q\bar{h}) - e^s q\bar{h}\sinh(q\bar{h})
\right ] \nonumber\\
&& + \left ( \frac{1+\nu^f_0}{1-\nu^f_0} \right ) \eta
\left ( \partial^2_z\hat{W} -\hat{\phi} \right )_{z=\bar{h}}
\left [ -b^s e^f \cosh(q\bar{h}) \right. \nonumber\\
&& \left. \left. -e^s b^f \sinh(q\bar{h}) +e^s q\bar{h}
\cosh(q\bar{h}) + (-b^s +\rho c^s) q\bar{h}\sinh(q\bar{h})
\right ] \right \},
\label{eq-a_xy}
\end{eqnarray}
\begin{eqnarray}
&iq_x\beta_x + iq_y\beta_y = & A^{-1} \left \{
q\bar{u}\hat{\zeta} \left [ -(1+\rho b^s) e^f b^f
+ (1+2\rho b^s -\rho^2 c^s)q^2\bar{h}^2 \right. \right. \nonumber\\
&& \left. +\rho e^s e^f (q\bar{h} +\sinh(q\bar{h})\cosh(q\bar{h}))
+(-\rho b^s e^f +\rho^2 c^s e^f)\sinh^2(q\bar{h}) \right ]
\nonumber\\
&& + \left [ q\bar{u}\hat{h} + \left ( \frac{1+\nu^f_0}
{1-\nu^f_0} \right ) \eta q \left ( \partial_z\hat{W}
\right )_{z=\bar{h}} \right ]  \left [ e^f (c^f +\rho b^s b^f)
\cosh(q\bar{h}) \right. \nonumber\\
&& \left. + \rho e^s {e^f}^2\sinh(q\bar{h})
-\rho e^s e^f q\bar{h}\cosh(q\bar{h}) -(c^f+2\rho b^s p^f
-\rho^2 c^s b^f) q\bar{h}\sinh(q\bar{h}) \right ] \nonumber\\
&& + \left ( \frac{1+\nu^f_0}{1-\nu^f_0} \right ) \eta
\left ( \partial^2_z\hat{W} -\hat{\phi} \right )_{z=\bar{h}}
\left [ -\rho e^s e^f b^f \cosh(q\bar{h}) \right. \nonumber\\
&& + (-c^f b^f +\rho b^s (2-4{\nu^f_0}^2) -\rho^2 c^s b^f) 
\sinh(q\bar{h}) \nonumber\\
&& \left. \left. +(c^f + 2\rho b^s p^f -\rho^2 c^s b^f)
q\bar{h}\cosh(q\bar{h}) +\rho e^s e^f q\bar{h}\sinh(q\bar{h})
\right ] \right \},
\label{eq-b_xy}
\end{eqnarray}
\begin{eqnarray}
q_y \alpha_x &=& q_x \alpha_y, \nonumber\\
q_y \beta_x &=& q_x \beta_y,
\label{eq-ab}
\end{eqnarray}
\begin{eqnarray}
& q \alpha_z = & A^{-1} \left \{
q\bar{u}\hat{\zeta} \left [ -(1+\rho b^s) {e^f}^2 
-(1+2\rho b^s -\rho^2 c^s)q^2\bar{h}^2 \right. \right. \nonumber\\
&& \left. -\rho e^s e^f (q\bar{h} +\sinh(q\bar{h})\cosh(q\bar{h}))
- (c^f +2\rho b^s p^f -\rho^2 c^s b^f) \sinh^2(q\bar{h}) 
\right ]  \nonumber\\
&& + \left [ q\bar{u}\hat{h} + \left ( \frac{1+\nu^f_0}
{1-\nu^f_0} \right ) \eta q \left ( \partial_z\hat{W}
\right )_{z=\bar{h}} \right ] \rho e^f
\left [ b^s e^f \cosh(q\bar{h}) \right. \nonumber\\
&& \left. +e^s b^f \sinh(q\bar{h}) +e^s q\bar{h}\cosh(q\bar{h})
+ (-b^s +\rho c^s) q\bar{h} \sinh(q\bar{h}) \right ]
 \nonumber\\
&& + \left ( \frac{1+\nu^f_0}{1-\nu^f_0} \right ) \eta
\left ( \partial^2_z\hat{W} -\hat{\phi} \right )_{z=\bar{h}}
\rho e^f \left [ -e^s e^f \cosh(q\bar{h}) \right. \nonumber\\
&& \left. \left. -(b^s b^f +c^s \rho)\sinh(q\bar{h}) + 
(b^s -c^s \rho)q\bar{h}\cosh(q\bar{h}) -e^s q\bar{h} \sinh(q\bar{h}) 
\right ] \right \},
\label{eq-a_z}
\end{eqnarray}
\begin{eqnarray}
& q\beta_z = & A^{-1} \left \{
q\bar{u}\hat{\zeta} \left [ -\rho e^s e^f b^f
- (c^f +2\rho b^s p^f -\rho^2 c^s b^f)(q\bar{h}
-\sinh(q\bar{h})\cosh(q\bar{h})) 
\right . \right. \nonumber\\
&& \left. +\rho e^s e^f \sinh^2(q\bar{h}) \right ]
+ \left [ q\bar{u}\hat{h} + \left ( \frac{1+\nu^f_0}
{1-\nu^f_0} \right ) \eta q \left ( \partial_z\hat{W}
\right )_{z=\bar{h}} \right ] \nonumber\\
&& \times \left [ \rho e^s e^f b^f \cosh(q\bar{h}) 
+ (c^f b^f -\rho b^s (2-4{\nu^f_0}^2) +\rho^2 c^s b^f) 
\sinh(q\bar{h}) \right. \nonumber\\
&& \left. +(c^f + 2\rho b^s p^f -\rho^2 c^s b^f)
q\bar{h}\cosh(q\bar{h}) +\rho e^s e^f q\bar{h}\sinh(q\bar{h})
\right ] \nonumber\\
&& + \left ( \frac{1+\nu^f_0}{1-\nu^f_0} \right ) \eta
\left ( \partial^2_z\hat{W} -\hat{\phi} \right )_{z=\bar{h}}
\left [ -e^f (c^f +\rho b^s b^f)\cosh(q\bar{h})
- \rho e^s {e^f}^2\sinh(q\bar{h}) \right. \nonumber\\
&& \left. \left. 
-\rho e^s e^f q\bar{h}\cosh(q\bar{h}) -(c^f+2\rho b^s p^f
-\rho^2 c^s b^f) q\bar{h}\sinh(q\bar{h}) \right ] \right \},
\label{eq-b_z}
\end{eqnarray}
and
\begin{eqnarray}
& C = & A^{-1} \left \{
q\bar{u}\hat{\zeta} \left [ -\rho e^s e^f
-(1+2\rho b^s -\rho^2 c^s)(q\bar{h}-
\sinh(q\bar{h})\cosh(q\bar{h})) \right ] \right. \nonumber\\
&& + \left [ q\bar{u}\hat{h} + \left ( \frac{1+\nu^f_0}
{1-\nu^f_0} \right ) \eta q \left ( \partial_z\hat{W}
\right )_{z=\bar{h}} \right ]
\left [ \rho e^s e^f \cosh(q\bar{h}) \right. \nonumber\\
&& \left. +(b^f -2\rho b^s \nu^f_0 +\rho^2 c^s)\sinh(q\bar{h})
+ (1 +2\rho b^s -\rho^2 c^s) q\bar{h}\cosh(q\bar{h})
\right ] \nonumber\\
&& + \left ( \frac{1+\nu^f_0}{1-\nu^f_0} \right ) \eta
\left ( \partial^2_z\hat{W} -\hat{\phi} \right )_{z=\bar{h}}
\left [ -e^f (1+\rho b^s) \cosh(q\bar{h}) \right. \nonumber\\
&& \left. \left. -\rho e^s e^f \sinh(q\bar{h})
-(1+2\rho b^s -\rho^2 c^s) q\bar{h} \sinh(q\bar{h}) 
\right ] \right \},
\label{eq-C_}
\end{eqnarray}
\begin{eqnarray}
& D = & A^{-1} \left \{
q\bar{u}\hat{\zeta} \left [ -e^f (1+\rho b^s)
-(1+2\rho b^s - \rho^2 c^s) \sinh^2(q\bar{h}) 
\right ] \right. \nonumber\\
&& + \left [ q\bar{u}\hat{h} + \left ( \frac{1+\nu^f_0}
{1-\nu^f_0} \right ) \eta q \left ( \partial_z\hat{W}
\right )_{z=\bar{h}} \right ]
\left [ e^f (1+\rho b^s) \cosh(q\bar{h}) \right. \nonumber\\
&& \left. +\rho e^s e^f \sinh(q\bar{h})
-(1+2\rho b^s -\rho^2 c^s) q\bar{h} \sinh(q\bar{h}) 
\right ] \nonumber\\
&& + \left ( \frac{1+\nu^f_0}{1-\nu^f_0} \right ) \eta
\left ( \partial^2_z\hat{W} -\hat{\phi} \right )_{z=\bar{h}}
\left [ -\rho e^s e^f \cosh(q\bar{h}) \right. \nonumber\\
&& \left. \left. -(b^f -2\rho b^s \nu^f_0 +\rho^2 c^s)
\sinh(q\bar{h}) + (1 +2\rho b^s -\rho^2 c^s) q\bar{h}
\cosh(q\bar{h}) \right ] \right \},
\label{eq-D_}
\end{eqnarray}
for the film solution, as well as
\begin{eqnarray}
u_x^0 &=& \alpha_x, \nonumber\\
u_y^0 &=& \alpha_y, \nonumber\\
u_z^0 &=& \bar{u} \hat{\zeta} + \alpha_z,
\label{eq-u^0}
\end{eqnarray}
and
\begin{equation}
B=\left ( q\bar{u}\hat{\zeta}
+ iq_x\alpha_x + iq_y\alpha_y +q\alpha_z \right ) /c^s,
\label{eq-B_}
\end{equation}
for the substrate solution. In equations (\ref{eq-a_xy})--%
(\ref{eq-D_}),
\begin{eqnarray}
& A = & {e^f}^2 + (1+2\rho b^s -\rho^2 c^s) q^2 \bar{h}^2
\nonumber\\
&& + 2\rho e^s e^f \sinh(q\bar{h})\cosh(q\bar{h})
+ (c^f +2\rho b^s b^f +\rho^2 c^s) \sinh^2(q\bar{h}),
\label{eq-A}
\end{eqnarray}
where $\rho$ is the relative film-substrate stiffness:
\begin{equation}
\rho = \mu^f_0 / \mu^s,
\label{eq-rho}
\end{equation}
and the elastic parameters $e^{f(s)}$, $b^{f(s)}$, $c^{f(s)}$,
and $p^{f(s)}$ are expressed by
\begin{eqnarray}
e^f = 2(1-\nu^f_0) \qquad &{\rm and}& \qquad
e^s = 2(1-\nu^s), \nonumber\\
b^f = 1-2\nu^f_0 \qquad &{\rm and}& \qquad
b^s = 1-2\nu^s, \nonumber\\
c^f = 3-4\nu^f_0 \qquad &{\rm and}& \qquad
c^s = 3-4\nu^s, \nonumber\\
p^f = 2-3\nu^f_0 \qquad &{\rm and}& \qquad
p^s = 2-3\nu^s.
\label{eq-ebcp}
\end{eqnarray}

To check whether these results are correct, we compare them with
the known solutions by taking appropriate limits. For the 
single-component system with planar film-substrate interface and
different elastic constants between film and substrate, Spencer et 
al. \cite{spencer4} have given the corresponding solution, which
can be recovered by setting $\hat{\zeta}=0$ as well as all the
terms related to $\hat{\phi}$ and $\hat{W}$ to be zero in the
above formulae (\ref{eq-a_xy})--(\ref{eq-A}). While for binary
alloy system ($\hat{\phi}\neq 0$) but with identical elastic
constants for film and substrate as well as flat interface,
the solution can be obtained by setting $\rho=1$, $\nu^f_0
=\nu^s=\nu_0$, $e^f=e^s$, $b^f=b^s$, $c^f=c^s$, $p^f=p^s$,
and $\hat{\zeta}=0$ in the above equations. The corresponding
results are presented in Eqs. (\ref{eq-u^f})--(\ref{eq-C})
of Sec. \ref{sec:III}, and consistent with that of previous
work. \cite{leonard3}

\section{\label{appendB}Elastic free energy in Fourier expansion}

Using the Fourier expansion form (\ref{eq-expan}), the elastic
free energy functional (\ref{eq-F_el-Eu}) can be expanded as
\begin{equation}
{\cal F}^f_{\rm el} = \int_0^{h} d^3r \left [
\bar{\cal E}^f + \sum\limits_{\bf q} e^{i(q_x x +q_y y)}
\hat{\cal E}^f (q,z,t) \right ] + \tilde{\cal F}^f_{\rm el},
\label{eq-F_el-q}
\end{equation}
with the zeroth order quantity
\begin{equation}
\bar{\cal E}^f = \frac{E_0}{1-\nu_0} \epsilon^2,
\label{eq-F0}
\end{equation}
which has no contribution to dynamic equations (\ref{eq-h}) and
(\ref{eq-phi}), and the first order term 
\begin{equation}
\hat{\cal E}^f = \frac{E_0}{1-\nu_0} \epsilon \left [
-\hat{u}^f_{ll} + \hat{u}^f_{zz} + \left ( 2\eta 
+\frac{2E_1^* - (1+\nu_0)\mu_1^*}{1-\nu_0} \epsilon \right )
\hat{\phi} \right ],
\label{eq-F1}
\end{equation}
which is used in evaluating the linear form of Eq. (\ref{eq-h})
for the surface height evolution. For the purpose of performing
the linear analysis on Eq. (\ref{eq-phi}) for the evolution of
$\phi$, we should use the second order elastic free energy
$\tilde{\cal F}^f_{\rm el}$:
\begin{eqnarray}
& \tilde{\cal F}^f_{\rm el} =& \sum\limits_{\bf q} \int^{\bar{h}}_{0} 
dz \left \{ \frac{E_0}{1-2\nu_0} \left [ \left ( \frac{3}{2}
\eta^2 + \frac{2(3E_1^*-2(1+\nu_0)\mu_1^*)}{1-\nu_0}\epsilon\eta
\right ) \hat{\phi}({\bf q}) \hat{\phi}(-{\bf q}) 
\right.\right. \nonumber\\
&& -\left. \left ( \eta + \frac{2E_1^*-(1+2\nu_0)\mu_1^*}{1-\nu_0}
\epsilon \right ) \hat{\phi}({\bf q}) \hat{u}_{ll}^f(-{\bf q})
+\frac{\nu_0}{2(1+\nu_0)} \hat{u}_{ll}^f({\bf q})
\hat{u}_{ll}^f(-{\bf q}) \right ] \nonumber\\
&& +\left. \frac{E_0}{1-\nu_0}\mu_1^* \epsilon \hat{\phi}({\bf q}) 
\hat{u}_{zz}^f(-{\bf q}) +\frac{E_0}{2(1+\nu_0)}
\hat{u}_{ij}^f({\bf q}) \hat{u}_{ij}^f(-{\bf q}) \right \},
\label{eq-F2}
\end{eqnarray}
which is different from the result in Ref. \onlinecite{leonard3}
where surface-bulk coupling of the film is neglected and the
second order elastic energy is evaluated at top surface $z=\bar{h}$.
If we calculate the functional differentiation $\delta {\cal F}^f_{el}
/ \delta \phi$ in Fourier space (that is, $\delta {\cal F}^f_{el} /
\delta \hat{\phi}(-{\bf q},z,t)$) using Eq. (\ref{eq-F2}), we can
also obtain the formula (\ref{eq-deltaF-q}) of Sec. \ref{sec:IV}
derived through real space procedure.

The expressions for first order elastic energy density (\ref{eq-F1}) 
evaluated at the surface and the functional differentiation Eq. 
(\ref{eq-deltaF-q}) in terms of $\hat{h}$ and $\hat{\phi}_s$ can be 
obtained by using the strain tensor formula (\ref{eq-u_lz}) of 
Appendix \ref{appendA} and substituting the form of $\hat{W}$ 
derived in Sec. \ref{sec:IV}, \textit{i.e.} $\hat{W}=v^2 \hat{\phi} 
/ (\Omega^2-q^2v^2)$, into the solution of mechanical equilibrium 
equation (\ref{eq-u^f})--(\ref{eq-C}) (with the approximation 
$\hat{\phi}|_{z=\bar{h}} = \hat{\phi}_s$ to first order). They 
are given by
\begin{equation}
\hat{\cal E}^f|_s = \frac{E_0}{1-\nu_0} \epsilon \left [
-2(1+\nu_0) \epsilon q \hat{h} + \left ( 2\eta
\frac{\Omega-\nu_0 qv}{\Omega+qv} + \frac{2E_1^*-(1+\nu_0)\mu_1^*}
{1-\nu_0} \epsilon \right ) \hat{\phi}_s \right ],
\label{eq-F1-s}
\end{equation}
and
\begin{eqnarray}
\left. \left ( \nabla^2\frac{\delta {\cal F}_{\rm el}}
{\delta \phi} \right )_q \right |_s &=& q^2 \frac{2E_0}{1-\nu_0} 
\left \{ \left ( \frac{1+\nu_0}{1-\nu_0} \right ) q 
\left [ (1-2\nu_0)\epsilon \eta + 
(2E_1^*-(1+\nu_0)\mu_1^*)\epsilon^2 \right ] \hat{h} 
\right. \nonumber\\
&-& \left [ \eta^2 + \frac{8E_1^*-5(1+\nu_0)\mu_1^*}
{2(1-\nu_0)}\epsilon \eta \right. \nonumber\\
&-& \left. \left. \left (\frac{1+\nu_0}{1-\nu_0} \right )
\left [ (1-2\nu_0)\eta^2 + (2E_1^*-(1+\nu_0)\mu_1^*)
\epsilon \eta \right ] \frac{qv}{\Omega+qv} \right ]
\hat{\phi}_s \right \},
\label{eq-deltaF}
\end{eqnarray}
respectively. Note that for the case of composition independent
elastic moduli ($E_1^*=\mu_1^*=0$), Eq. (\ref{eq-F1-s}) yields 
a result for first order elastic energy density which is the same 
as that in Ref. \onlinecite{leonard3}. The differences occur in Eq.
(\ref{eq-deltaF}) due to differences in $\tilde{\cal F}^f_{\rm el}$
as well as different model and derivation procedure: 
In the previous work of Ref. \onlinecite{leonard3}, the second 
order elastic energy (\ref{eq-F2}) is evaluated at surface 
$z=\bar{h}$ first and expressed in terms of $\hat{h}$ and 
$\hat{\phi}|_{z=\bar{h}}$, and then the functional differentiation
is carried out directly on surface $\phi$ with an approximation
that $\delta\tilde{\cal F}^f_{\rm el}/\delta \hat{\phi}_s
\sim \delta\tilde{\cal F}^f_{\rm el}/\delta \hat{\phi}|_{z=\bar{h}}$.

\newpage

\begin{thebibliography}{}
\bibitem{nonzero-zeta}In general the roughness of the film-substrate 
interface may necessitate its location at $z=\zeta (x,y)$ instead of 
at $z=0$. We have investigated whether a nonzero $\zeta$ contributes 
to any instability at the top surface $z=h(x,y)$, using the general
solution of mechanical equilibrium equation in Appendix \ref{appendA}. 
Our conclusion is that an undulation at the film-substrate interface 
results in an exponentially damped perturbation at top surface as the 
film grows and is thus benign.
\bibitem{note-gamma} We have also studied the case where $\gamma$ may
have a linear composition dependence and found that it has no extra 
contribution to the results of linear stability analysis.
\bibitem{note-dev} More precisely, the functional differentiation
of free energy functional ${\cal F}$ with respect to $\phi$ generates 
an extra contribution proportional to $\delta (z-h)$ (mostly caused
by $u_{ij}^{hom}$), corresponding to an additional boundary condition 
for compositional variable at surface $z=h$ which is not relevant
in linear analysis.
\end{thebibliography}
\end{document}